\documentclass[prd,twocolumn,showpacs,showkeys]{revtex4}
\usepackage{graphicx}
\usepackage{dcolumn}
\usepackage{bm}
\usepackage{amssymb}
\usepackage{comment}
\usepackage{mathrsfs}
\usepackage{amsmath}

\begin{document}
\title{Photon Magnetic Moment and Vacuum Magnetization \\
in an Asymptotically Large Magnetic Field}
\author{Selym \surname{Villalba-Ch\'avez}}
\affiliation{Institute of Physics,  University of Graz, Universit\"atsplatz 5, 8010, Graz,
Austria.\\ on leave at Universidad Complutense de Madrid, Spain.}
\date{\today}
\begin{abstract}
We consider the effect of the photon radiative correction on the
vacuum energy in a superstrong magnetic field. The notion of a  photon
anomalous magnetic  moment is analyzed and its connection with  the quasiparticle character of the electromagnetic radiation is established. In
the infrared domain the magnetic moment turns out to be a vector
with two orthogonal components in correspondence with the
cylindrical symmetry imposed by the external field. The possibility
of defining such quantity in the high energy limit is  studied as well.
Its existence suggests that the electromagnetic radiation is a
source of magnetization to the whole vacuum and thus its
electron-positron zero-point energy  is slightly modified. The
corresponding contribution to the vacuum magnetization density is
determined by considering the individual contribution of each vacuum
polarization eigenmode in the Euler-Heisenberg Lagrangian. A
paramagnetic response is found in one  of them, whereas the
remaining ones are diamagnetic. Additional issues concerning  the
transverse pressures are analyzed.
\end{abstract}



\maketitle
\section{Introduction}

Large magnetic fields $\vert\textbf{B}\vert\gg \rm B_c,$ $\rm B_c=m^2/e=4.42\cdot 10^{13}
G$ ($\rm m$ and $\rm e$ are the electron mass  and charge,
respectively) in the surface of stellar objects identified as
neutron stars \cite{Manchester,Kouveliotou,Bloom} might provide
physical scenarios where quantum processes predicted in such a
regime could  become relevant for astrophysics and cosmology. According to  quantum
electrodynamics(QED) in strong background fields, the most important effects are likely to be
pair creation \cite{sauter,euler,Schwinger}, photon splitting \cite{adler,Adler:1971wn}  and photon capture
\cite{shabad1,shabad2,shabad3v,shabad3,shabad4}. The last two essentially depend  
on the drastic departure of the photon dispersion relation from the light cone curve, due to  the vacuum
polarization tensor $\Pi_{\mu\nu}$ which depends on both the Landau levels
of  virtual electron-positron pair, as well as on the external magnetic field.
As a result, the issue of light propagation in empty space, in the
presence of $\bf B,$ is  similar to the dispersion of
light in an anisotropic ``medium'',  with the  preferred direction
corresponding to the external field axis.  The  phenomenological aspects associated with this issue  has been studied for a long time and recently,  the  effects  of the vacuum polarization tensor on the Coulomb potential in superstrong magnetic field  have been considered as well \cite{shabad5,shabad6,Sadooghi:2007ys}.  However, the  problem concerning  the  magnetism carried by a photon  has not attracted
sufficient attention,  excepting   \cite{hugoe}, where  an  photon anomalous magnetic moment has  been  pointed out. The  authors of this reference  attempted to derive  this  quantity in  two different regimes of the vacuum polarization tensor.  On the one
hand for low  energies in weak fields ($\vert\bf B\vert\ll\rm B_c$)
and on the other hand (originally studied in Ref. \cite{selym})
near the first pair creation threshold and for a moderate fields
$(\vert\bf B\vert\sim \rm B_c)$. However, it was  ignored that the
photon magnetic moment $\boldsymbol{m}_\gamma$ actually  is  a vector defined as  the coefficient of $\bf B$ when the energy is linearly approximated  in term of the external magnetic field. In consequence, neither any  discussion about the connection  between this  quantity and  its angular momentum was  presented nor any   comment  about its  precession around the external field axis was made. 

In the  present  work we reveal the asymptotic conditions where the 
concept of a photon anomalous magnetic moment  may be adequate. We will analyze the cases
of superstrong magnetic fields ($\vert \bf B\vert\gg \rm B_c$) for
both the low and the high energy limit of the vacuum polarization tensor. In
these  domains,  one eigenvalue of $\Pi_{\mu\nu}$ depends linearly on the external
field.   In  consequence, the Maxwell equations in the ``medium''  seem  to describe a  massless particle with a  magnetic moment. In this case $\boldsymbol{m}_\gamma$   may become a 
characteristic quantity since  the external field generates a Lorentz symmetry break down  which induces the  nonconservation  of the photon helicity. We shall see that this notion equips us with an intuitive tool  to understand, in a phenomenological way, the quasiparticle behavior of a  photon propagation mode in a highly  magnetized vacuum.  

The second purpose of this work is to show that virtual photons are a source of magnetization to the whole vacuum. We will  address the question in which way the virtual electromagnetic radiation   contributes to
a measurement of the vacuum magnetization and therefore to increase
the external field strength. This aspect might be  important for 
astrophysics since  the  origin and  evolution of  magnetic fields in 
compact stellar objects  remains  poorly understood  \cite{Uzdensky:2009cg}.  
Some investigations in this area  provide theoretical  evidences that $\vert\textbf{B}\vert$ is self-consistent due to the  Bose-Einstein condensation  of charged and 
neutral boson gases in a superstrong magnetic field \cite{Chaichian:1999gd,Martinez:2003dz,PerezRojas:2004ip,PerezRojas:2004in}. 
In  this context, the nonlinear QED-vacuum  possesses the properties of a 
paramagnetic medium and seems to play an important role within the 
process of magnetization in the stars.  Its properties  have been
studied also in  \cite{hugo3,hugo4,Elizabeth1,Elizabeth2} for weak
($\vert\bf B\vert\ll \rm B_c$) and  moderate fields  ($\vert\bf B\vert\sim \rm B_c$) in one-loop
approximation of the Euler-Heisenberg Lagrangian \cite{euler}. New
corrections   emerge by considering  the two-loop term of this
effective Lagrangian \cite{ritus,ritus1,dittrich1,schubert,kors}
which contains  the contribution of virtual photons created and
annihilated spontaneously in the vacuum and  interacting with the
external field through the vacuum polarization tensor.  

The  two-loop term  of the Euler-Heisenberg Lagragian  was computed many years ago  by  Ritus \cite{ritus,ritus1}.   A few years latter, Dittrich and Reuter \cite{dittrich1} obtained  a  simpler integral
representation of this term and  showed that 
their  results agreed with those determined by Ritus in the strong magnetic
field limit.  In the last few years, it  has been recalculated by
several authors  using the  worldline formalism \cite{schubert,kors}
and it has been extended to the case of finite temperature as well \cite{gies}.
Nevertheless, in all these works it is really cumbersome to discern
the individual contributions given by each virtual photon propagation
mode to the Euler-Heisenberg Lagrangian which should  allow to determine the global magnetic character
generated by each of them.  In this paper we compute these
contributions separately. In addition we  derive   the individual vacuum magnetizations
provided  by each virtual photon propagation mode for very large
magnetic fields  $(\vert\bf B\vert\gg\rm B_c)$. Whilst 
the complete two-loop contribution  is purely paramagnetic we find a
diamagnetic response  coming from  that $\Pi_{\mu\nu}-$eigenvalue which is 
used to obtain the  photon anomalous magnetic moment. 

In order to expose  our results we have structured our paper  as follows. 
In  Sect. II we recall some basic features of 
photon propagation in an external magnetic field.   Some aspects of the Noether 
currents  associated with  the  spacetime symmetries  in presence of
an  $\bf B$  are  discussed  in Sect. III. The corresponding calculations  are carried out  in a
basis that diagonalizes the polarization tensor.  The  posibility to define a photon
anomalous magnetic moment in  an asymptotically large magnetic field is studied in Sect. IV.  We shall  show that   $\boldsymbol{m}_\gamma$ turns out to be a vector with two orthogonal components in correspondence with the cylindrical symmetry imposed by the external field.  Moreover, we will see that a nonvanishing torque exerted by $\textbf{B}$ might generates a  precession of $\boldsymbol{m}_\gamma$  which   is a
manifestation of the reduction  of the rotation symmetry and,
consequently  Lorentz symmetry.   In Sect. V we perform the calculation of  the  two-loop  contributions to  the
Euler-Heisenberg Lagrangian given by each virtual mode. From them we
obtain  the modified vacuum energy for very large
magnetic fields in Sect. VI. The corresponding vacuum magnetizations and magnetic
susceptibilities are analyzed as well as the transversal pressure.
There is additional discussion  given in the conclusions while essential steps
of many calculations have been deferred to the appendices.

\section{General remarks}
\subsection{Symmetry reduction and diagonal decomposition of vacuum polarization tensor}

The standard relativistic description of a photon reflects the
underlying symmetry  of the Poincar\'e group
$(\rm{ISO}(3,1))$. Its  irreducible representations are  characterized by two observable degrees of freedom,
corresponding to the helicity values $\lambda=\bf J\cdot k/\vert\bf
k\vert=\pm1.$  Here, $\bf J$ encodes the
generators  of the $3-$dimensional rotation group and $\bf k$ is the
photon momentum. In the presence of an external 
electromagnetic field, space-time is no longer isotropic and
Poincar\'e invariance breaks down. In a reference frame where the
field  is purely magnetic, the symmetry breaking   for neutral
particles has  the following pattern:
\begin{equation}
\rm{ISO}(3,1)+\textbf{B}\to\rm{ISO}(2)\times\rm{ISO}(1,1).\label{fundamentalgroupdecompo}
\end{equation}  The effective physical configuration of Minkowski space  manifests itself as the direct product of the  $2-$dimensional Euclidean group $\rm{ISO}(2)$ and the $(1+1)-$dimensional pseudoeuclidean group
$\rm{ISO}(1,1)$. The two resulting symmetry groups are
associated with the transversal and pseudoparallel planes with respect to  $\bf
B-$direction. The conserved quantities  of any uncharged relativistic quantum
field within this reference frame  are  related to the Casimir
invariants of this direct product of groups \cite{Bacry}.

For the electromagnetic  radiation this type symmetry reduction occurs   as
soon as radiative corrections are considered. In such a case an
observable photon interacts with the external magnetic field through
the virtual electron-positron ($e^{\mp}$) pair whose Green's
functions $\mathscr{G}(\rm x,x^\prime\vert \rm B)$ determine the
vacuum polarization  tensor $\Pi_{\mu\nu}(\rm x,x^\prime\vert
B)=-e^2\mathrm{Tr}\left[\gamma_\mu\mathscr{G}(x,x^\prime\vert \rm
B)\gamma_\nu\mathscr{G}(x^\prime,x\vert \rm B)\right].$ Consequently,
the photon behaves like a quasiparticle embodying  both radiation
and $e^\mp$ properties, the latter  quantized by the Landau levels.

The diagonalization of $\Pi_{\mu\nu}$ is  expressed as
\begin{equation}
\Pi_{\mu\nu}=\sum_{i=0}^4\varkappa_i(\rm z_1, \rm z_2, \Delta)\frac{\rm a_\mu^{(i)}\rm
a_{\nu}^{(i)}}{\left(\rm a^{(i)}\right)^2},\label{gstrpi}
\end{equation} with   its  renormalized eigenvalues $\varkappa_i$
depending  on  the scalars,  $\rm z_1=\rm kF^{*2}k/(2\Delta)$ and $\rm z_2=-\rm kF^2k/(2\Delta),$  which together with $\rm k^2=\rm z_1+\rm z_2$ form the Casimir invariants of the $\rm{ISO}(2)\times\rm{ISO}(1,1)$ Lie algebra.  Here $\rm F^{\mu\nu*}=1/2\epsilon^{\mu\nu\rho\sigma}\rm F_{\rho\sigma}$  represents the dual of $\rm F_{\mu\nu},$ whereas  $\Delta=1/4\rm F_{\mu\nu}F^{\mu\nu}=1/2\vert\textbf{B}\vert^2$ is one of the external field  invariants (the remaining one, $\Phi=1/4\rm F^{\mu\nu}F_{\mu\nu}^{*}$, vanishes identically). 

In one-loop approximation, the eigenvalues of $\Pi_{\mu\nu}$ read
\begin{eqnarray}
\begin{array}{c}\displaystyle
\varkappa_1=-\frac{1}{2}\rm k^2 \mathscr{I}_1,\\  \displaystyle
\varkappa_2=- \frac{1}{2}(\rm z_1\mathscr{I}_2+\rm z_2\mathscr{I}_1),\ \ \varkappa_3=-\frac{1}{2}(\rm z_1\mathscr{I}_1+\rm z_2\mathscr{I}_3), 
\end{array}\label{poleig1}
\end{eqnarray}
with
\begin{eqnarray}
\mathscr{I}_i&=&\frac{4\alpha}{\pi}\int_0^\infty\rm d\tau\rm e^{-\rm m^2\tau}\int_{0}^1 \rm d\eta\frac{\rm eB\sigma_i}{\sinh(\rm s)}\rm \exp\left\{-\rm z_2\frac{M(\rm s,\eta)}{\mathrm{eB}}\right.\nonumber\\&-&\left.\rm z_1\frac{N(\rm s,\eta)}{\mathrm{eB}}\right\}-\frac{4\alpha}{\pi}\int_0^\infty\frac{\rm d\tau}{\tau}\rm e^{-\rm m^2\tau}\int_{0}^1 \rm d\eta\frac{1-\eta^2}{4}.\label{poleig2}
\end{eqnarray} Here and in the following  $\rm s\equiv eB\tau,$
\begin{eqnarray}
\begin{array}{c}
\displaystyle \sigma_{1}(\rm s,\eta)=\frac{1}{4}\frac{\sinh(\rm s)\cosh(\rm s\eta)-\eta \sinh(\rm s \eta)\cosh(\rm s)}{\sinh(\rm s)},\\ \\
\displaystyle \sigma_2(\rm s,\eta)=\frac{1-\eta^2}{4}\cosh(\rm s),\ \ \sigma_3(\rm s,\eta)=\frac{\rm M(\rm s,\eta)}{\sinh(\rm s)},\\ \\ \displaystyle \rm M(\rm s,\eta)=\frac{\cosh(\rm s)-\cosh(\rm s\eta)}{2\sinh(\rm s)},\ \ \rm N(\rm s,\eta)=\frac{1-\eta^2}{4}\rm s.\end{array}\label{poleig4}
\end{eqnarray}

The four vector $\rm a^{(i)}_\mu$ in Eq. (\ref{gstrpi}) denotes the  corresponding
eigenvectors of $\Pi_{\mu\nu}$ \cite{batalin}:
\begin{eqnarray}
\begin{array}{c}
\displaystyle \rm
a^{(1)}_\mu=\frac{\rm k^2 F^2_{\mu\lambda}k^\lambda-(kF^2 k)k_\mu
}{\rm k^2 (-\rm kF^2k)^{1/2}}, \ \ \rm a^{(2)}_\mu =\frac{\rm
F^{*}_{\mu\lambda}k^\lambda}{(2\Delta)^{1/2}},\\  \displaystyle \rm
a^{(3)}_\mu=\frac{\rm F_{\mu \lambda}k^\lambda}{(-\rm kF^2k)^{1/2}} \
\ \ \ \mathrm{and} \ \ \ \ \rm a^{(4)}_\mu= \frac{\rm k^\mu}{\rm k^2}
\end{array}
\end{eqnarray} which  fulfill both the orthogonality condition:
$\rm a_{\sigma}^{(i)}a^{\sigma(j)}=\delta^{ij}\left(\rm
a^{(i)}\right)^2$ and  the completeness relation:
$\delta^{\mu}_{\quad\nu}=\sum_{i=1}^4\rm a^{\mu (i)}\rm
a_{\nu}^{(i)}/\left(a^{(i)}\right)^2.$

Owing  to the transversality
property $(\rm k^\mu\Pi_{\mu\nu}=0),$  the  eigenvalue corresponding to
the fourth eigenvector  vanishes identically $(\varkappa^{(4)}=0).$
Consequently, the photon propagator can be decomposed as
\begin{equation}
\mathfrak{D}_{\mu\nu}=\sum_{i=1}^3\frac{1}{\textrm{k}^2-\varkappa^{(i)}}\frac{\rm
a_\mu^{(i)}\rm a_{\nu}^{(i)}}{\left(\rm a^{(i)}\right)^2}-\frac{\zeta}{\rm k^2}\frac{\rm
k_\mu k_\nu}{\rm k^2}\label{photonpro}
\end{equation} with   $\zeta$ being the gauge parameter.

According to  Eq. (\ref{photonpro}) three nontrivial  dispersion relations  arise 
\begin{equation}
\rm k^2=\varkappa_i\left(\rm z_2,\rm z_1,\Delta\right) \
\ \mathrm{for} \ \  i=1,2,3  \label{egg}.
\end{equation} We should keep in mind that the general  structures of $\rm z_2$ and $\rm z_1$ are complicated (see Appendix \ref{hfgt}) since both of them depend on the relative $\textbf{B}-$orientation with respect to a reference frame. Substantial  simplifications are achieved  in  reference frames which are either at rest or moving parallel to the external field. In these cases $\rm z_2=\rm k_\perp^2$ and $\rm z_1=\rm k_\parallel^2-\omega^2.$ Here  ${\bf k}_{\perp}$ and ${\bf k}_{\parallel}$ are the components of $\textbf{k}$ perpendicular and along the external field respectively, with   $\rm k^2=k_\perp^2+k_\parallel^2-\omega^2$.   Note that, as a consequence,   $\varkappa_i$  depend on both   $\Delta=1/2(\rm B_x^2+B_y^2+B_z^2)$  and  on the  $\textbf{B}$-direction with respect to  our reference frame.

Solving Eq. (\ref{egg}) for $\rm k^2=\textbf{k}^2-\omega^2$ in terms of $\rm z_2$ yields
\begin{equation}
\omega_i^2=\textbf{k}^2+\mathfrak{m}_{i}^2(\rm z_2,\Delta),\label{gdl}
\end{equation} with the term $\mathfrak{m}_i$ arising  as a sort of dynamical mass. Note that $\mathfrak{m}_i$ vanishes for  $\rm z_2=0$ due to the gauge invariance condition $\varkappa_i(0,0,\Delta)=0.$ Obviously, the dispersion law given in  Eq. (\ref{gdl})  differs from the usual light cone equation. This difference increases   near the free pair creation thresholds  remarking  the quasiparticle feature of a photon in an external magnetic field. For more details we refer the reader to Ref. \cite{shabad1}.

By considering $\textrm{a}^{(i)}_\mu(\rm k)$ as the
electromagnetic four vector describing the eigenmodes, we obtain the
corresponding electric and magnetic fields of each mode
\begin{equation}
{\bf e}^{(i)}=i(\omega^{(i)}\textbf{a}^{(i)}-\textbf{k} \textrm{a}^{(i)}_0)\ \ \mathrm{and} \ \ {\bf
h}^{(i)}=-i\textbf{k}\times\textbf{a}^{(i)}. \label{electricpromag}
\end{equation}
Up to a non essential proportionality factor, they are explicitly given by:
\begin{eqnarray}
\begin{array}{c}
\textbf{e}^{(1)}=-i\textbf{n}_\perp\omega,\ \ \textbf{h}^{(1)}=i\textbf{k}_\parallel\times\textbf{n}_\perp,\\ 
\textbf{e}_\perp^{(2)}=i\textbf{k}_\perp \textrm{k}_{\parallel},\ \ \textbf{e}_\parallel^{(2)}=i\textbf{n}_{\parallel}(\textrm{k}_{
\parallel}^2-\omega^2),\\ \textbf{h}^{(2)}=i\omega(\textbf{k}_{\perp}\times\textbf{n}_\parallel),\\ 
\textbf{e}^{(3)}=i\omega (\textbf{n}_\perp\times
\textbf{n}_{\parallel}),\\\textbf{h}_\parallel^{(3)}=i\textbf{n}_{\parallel}\rm
k_\perp,\ \ \textbf{h}_\perp^{(3)}=-i\textbf{n}_\perp \textrm{k}_{\parallel}.
\end{array}\label{electric-magnetic-field}
\end{eqnarray}
Here,   $\bf n_\parallel=
k_\parallel/\vert k_\parallel\vert$ and $\bf n_\perp=
k_\perp/\vert k_\perp\vert$ are the unit vectors associated with the parallel and perpendicular  direction with respect $\bf B$.

\section{Noether currents of the radiation Field in presence of $\vec{{\bf B}}$ \label{sectspin}}

\subsection{The Poynting vector and the physical propagation modes}

Whenever the Minkowski space  is   translations  invariant,  the associated  Noether current of the electromagnetic  radiation ($\textrm{A}_\mu(\rm x)$) provides a conserved   stress-energy tensor:    
\begin{equation}
\rm T^{\mu\nu}=\mathfrak{F}^{\mu\lambda}\mathfrak{F}_{\ \ \lambda}^{\nu}-\frac{1}{4}\eta^{\mu\nu}\mathfrak{F}_{\sigma\lambda}\mathfrak{F}^{\sigma\lambda}. \label{moemntumeberguf}
\end{equation} Here $\mathfrak{F}_{\mu\nu}=\partial_\mu
\textrm{A}_\nu-\partial_\nu \textrm{A}_\mu$ is the electromagnetic tensor,  whereas  the metric tensor $\eta^{\mu\nu}$ has  signature $+++-$ with  $\eta^{11}=\eta^{22}=\eta^{33}=-\eta^{00}=1.$ In this context the momentum density  is defined by $\mathfrak{P}^\mu\equiv \rm T^{0\mu}.$  Note that the symmetry reduction by the external magnetic field does not alter the translational group  involved within $\rm ISO(3,1)$. Therefore,  all $\mathfrak{P}^\mu-$components are conserved.

The spatial part of $\mathfrak{P}$  is the Poynting vector density
\begin{equation}
\boldsymbol{\mathcal{P}}=\textbf{E}\times \textbf{H}.
\end{equation}  For each eigenmode it reads  $\boldsymbol{\mathfrak{p}}^{(\mathrm{i})}=\textbf{e}^{(\mathrm{i})}\times\textbf{h}^{(\mathrm{i})}$. In particular
\begin{equation}
\begin{array}{c}
\boldsymbol{\mathfrak{p}}^{(\mathrm{1})}=\omega \textbf{k}_\parallel,\ \
 \boldsymbol{\mathfrak{p}}^{(\mathrm{3})}=\omega \textbf{k}, \\ \\
\boldsymbol{\mathfrak{p}}_\parallel^{(\mathrm{2})}=\omega \textrm{k}_\perp^2 \textbf{k}_\parallel,\ \ \boldsymbol{\mathfrak{p}}_\perp^{(\mathrm{2})}=\omega(\omega^2-\textrm{k}_\parallel^2)\textbf{k}_\perp.
\end{array}
\end{equation}

Different  photon degrees of freedom contribute in the presence of ${\bf B},$ depending  on  the direction of wave
propagation: for a pure longitudinal propagation $\bf k\parallel B,$ the Poynting vector  $\boldsymbol{\mathfrak{p}}^{(2)}=0.$ As a consequence, eigenmode $2$ is a  pure longitudinal  non physical electric wave and does not carry energy. On the other hand, the first and third mode have  well-defined Poynting vectors  along the external field $\boldsymbol{\mathfrak{p}}^{(\mathrm{1})}=\boldsymbol{\mathfrak{p}}^{(\mathrm{3})}.$ In this case,  each set  $\left\{\textbf{e}^{(1)},\textbf{h}^{(1)}, \boldsymbol{\mathfrak{p}}^{(\mathrm{1})}\right\}$ and $\left\{\textbf{e}^{(3)},\textbf{h}^{(3)}, \boldsymbol{\mathfrak{p}}^{(\mathrm{3})}\right\}$  forms an orthogonal set of vectors and represent waves  polarized in the transverse plane to $\textbf{B}.$  Consequently, for pure parallel propagation  $\rm a_{\mu}^{(1)}$ and $\rm a_{\mu}^{(3)}$  represent  physical waves. 

Now, if the photon propagation  involves a nonvanishing  transversal momentum component $\rm k_\perp\neq0,$ we are allowed to  perform the analysis in a Lorentz frame that  does not change the value $\rm k_\perp$,  but gives $\rm k_\parallel=0$ and does not introduce an external electric field. As a consequence, the  energy flux of the first eigenmode $\boldsymbol{\mathfrak{p}}^{(1)}=0$ and   becomes purely electric longitudinal and a  non physical mode. In the same context each set  $\left\{\textbf{e}^{(2)},\textbf{h}^{(2)}, \boldsymbol{\mathfrak{p}}^{(\mathrm{2})}\right\}$ and $\left\{\textbf{e}^{(3)},\textbf{h}^{(3)}, \boldsymbol{\mathfrak{p}}^{(\mathrm{3})}\right\}$  forms an orthogonal set of vectors and represent waves  polarized in the transverse plane to $\textbf{B}.$ Hence, for  a photon whose three-momentum  is directed at any nonzero angle with the external magnetic field, the  two orthogonal polarization states $\rm a_\mu^{(2)}$ and $\rm a_\mu^{(3)}$ propagate. Note that mode $3$ represents a physical wave  independent of the direction of propagation.

\subsection{The spin and boost  of the eigenwaves\label{spingenerallrofjdffk}}

The Noether current associated  with rotational
invariance is usually split  into two pieces:
orbital angular momentum density and intrinsic spin density.
For an electromagnetic field,
the latter can be derived from the third rank tensor
\begin{equation}
\mathcal{S}^\mu_{\alpha\beta}(x)=\frac{\partial\mathscr{L}_0}{\partial(\partial_\mu
\textrm{A}_\nu(x))}\left(\mathscr{J}_{\alpha\beta}\right)_\nu^\sigma
\textrm{A}_\sigma(x)\label{sp}
\end{equation} with  $\mathscr{L}_0=-\frac{1}{4}\mathfrak{F}_{\mu\nu}\mathfrak{F}^{\mu\nu}$ being
the free part of the QED-Lagrangian. Here 
$\left(\mathscr{J}_{\alpha\beta}\right)^\sigma_\nu=-i\left(\delta^\sigma_\alpha\eta_{\beta\nu}
-\delta^\sigma_\beta\eta_{\alpha\nu}\right)$ denotes  the four-dimensional
representation of the Lorentz generators. By considering the
previous definition we can express  Eq. (\ref{sp}) as
\begin{equation}
\mathcal{S}^{\mu}_{\alpha\beta}(x)=-i\left[\mathfrak{F}_{\alpha}^{\mu}\textrm{A}_{\beta}(x)-\mathfrak{F}_{\beta}^{\mu}\textrm{A}_{\alpha}(x)\right].\label{curentss}
\end{equation} 

We fix   $\mu=0$ in Eq. (\ref{curentss}) and take into account the spatial part of the
remaining tensor. Under such a condition the intrinsic angular
momentum density of the electromagnetic field  is reduced to
\begin{equation}
\mathcal{S}^{0}_{ij}=-i\left[\mathrm
{E}_{i}(x)\mathrm{A}_{j}(x)-\mathrm{E}_{j}(x)\mathrm{A}_{i}(x)\right],\label{spinvect}
\end{equation} with $\mathrm{E}^{i}(x)=\mathfrak{F}^{i0}$ being the  electric field. The spatial
 components defined from $\frac{1}{2}\epsilon^{ijk}\mathcal{S}^{0}_{ij}$ with
$\epsilon^{123}=1$ define  the  classical spin density  of the
electromagnetic field
\begin{equation}
\boldsymbol{\mathcal{S}}(\textbf {x},\textrm{t})=-i \textbf{E}(\textbf {x},\rm
t)\times\textbf{A}(\textbf{x},\rm t).\label{spinvectorial}
\end{equation}

For $\mu=\beta=0$ we obtain the corresponding  Noether current
related to  the boost transformations, which is given by 
$\boldsymbol{\mathfrak{K}}(\textbf {x},\textrm{t})=i\textbf{E}(\textbf
{x},\rm t)\textrm{A}_0(\textbf{x},\rm t).$

In order  to analyze the behavior of $\boldsymbol{\mathcal{S}}$ and $\boldsymbol{\mathfrak{K}}$  in presence  of
$\bf B$ we regard $\textbf{e}^{(i)}$  (see Eq. (\ref{electricpromag})) as the electric  field associated with  the fourth potential $(\textbf{a}^{(i)}).$ Up to a nonessential factor of proportionality  we have
\begin{eqnarray}
\boldsymbol{\mathfrak{s}}^{(i)}=-i\textbf{e}^{(i)}\times\textbf{a}^{(i)}=-i \textrm{a}_0^{(i)}\textbf{h}^{(i)}\ \ \mathrm{and}\ \
\boldsymbol{\kappa}^{(i)}=i\textrm{a}_0^{(i)}\textbf{e}^{(i)}.
\label{passss1}
\end{eqnarray}
Manifestly Eq. (\ref{passss1}) expresses the connection between  the
Noether  currents and the different polarization planes associated
with  each eigemode. In particular
\begin{eqnarray}
\begin{array}{c}
\displaystyle\boldsymbol{\mathfrak{s}}^{(1)}=\frac{\omega}{\rm k^2}(\textbf{k}_\parallel\times\textbf{k}_\perp),\ \ \boldsymbol{\kappa}^{(1)}=\frac{\omega^2}{\textrm{k}^2}\textbf{k}_\perp,\\ \\
\boldsymbol{\mathfrak{s}}^{(2)}=\omega(\textbf{k}_\parallel\times\textbf{k}_\perp),\\
\boldsymbol{\kappa}_\perp^{(2)}=\textbf{k}_\perp \rm
k_{\parallel}^2,\ \ \boldsymbol{\kappa}_\parallel^{(2)}=\textbf{k}_{\parallel}(\textrm{k}_{
\parallel}^2-\omega^2),\\ \\
\boldsymbol{\mathfrak{s}}^{(3)}=0,\ \
\boldsymbol{\kappa}^{(3)}=0.
\end{array}\label{idncja}
\end{eqnarray}

According to  the reduced Lorentz symmetry, \emph{i.e.}  Eq.
(\ref{fundamentalgroupdecompo}), only the parallel components
$\mathfrak{s}_\parallel^{(i)}$ and $\kappa_\parallel^{(i)}$ are
related to conserved quantities  and therefore, just the component
of the electric and magnetic field along $\mathbf{B}$ can generate
conserved charges. However,  for a   purely   parallel propagation
$({\bf k}_\perp=0)$,  this connection is rather obscure due to the
absence of $\mathbf{e}_\parallel^{(1,3)}$ and
$\mathbf{h}_\parallel^{(1,3)}.$ In this case the first and third
mode  may be combined to form a  circularly polarized transversal
wave which is   allowed by the  degeneracy property:
\begin{equation}
\varkappa_1(\rm
z_1,0,\mathfrak{F})=\varkappa_3(\rm z_1,0,\mathfrak{F}).
\end{equation} In  this context the photon is labeled by the helicity
$\lambda=\textbf{J}_\parallel\cdot\textbf{n}_\parallel$  which seems
to be a conserved quantity.  For nonvanishing  transversal
propagation $\textbf{k}_\perp\neq0,$  however, $\lambda$  stops
being a well-defined quantum number due to  the nonconserved
rotations transversal  to $\textbf{B}.$

\subsection{Connection between helicity and spin for pure parallel propagation}

In order to establish the connection between helicity and 
classical spin of the electromagnetic field  we express
$\mathcal{S}_{\alpha\beta}^\mu=\mathcal{S}_{\alpha\beta}^{\mu+}+\mathcal{S}_{\alpha\beta}^{\mu-}.$ Here
\begin{eqnarray}
\mathcal{S}_{\alpha\beta}^{\mu\pm}=-\frac{i}{2}\left\{\mathfrak{F}_{\alpha}^\mu\pm i\mathfrak{F}_{\alpha}^{*\mu}\right\}\textrm{A}_\beta-\frac{i}{2}\left\{\mathfrak{F}_{\beta}^\mu\pm i\mathfrak{F}_{\beta}^{*\mu}\right\}\textrm{A}_\alpha\label{passss}
\end{eqnarray}  and  $\mathfrak{F}_{\alpha}^{*\mu}$ is  the dual of $\mathfrak{F}_{\alpha}^{\mu}.$  Adopting  a  similar procedure to  those developed below Eq. (\ref{curentss})  the vectors read
\begin{equation}
\boldsymbol{\mathcal{S}}^{\pm}=-\frac{i}{2}\left\{\mathbf{E}\pm i\mathbf{H}\right\}\times\textbf{A}.\label{passsspm}
\end{equation} The complex fields  $\frac{1}{2}\left\{\mathbf{E}\pm i\mathbf{H}\right\}$  transform irreducibly under spin $(1,0)$ and $(0,1)$ representation of $\rm SO(3,1)\sim\rm SU(2)\times SU(2),$ respectively. These field combinations fulfill the free Maxwell  equation  for left- and right circularly polarized radiation
\begin{equation}
\boldsymbol{\nabla}\times (\textbf{E}\pm i\textbf{H})\mp i\frac{\partial}{\partial\textrm{t}}(\textbf{E}\pm i\textbf{H})=0 
\end{equation} corresponding to $\lambda=\mp1.$

For  propagation  purely parallel to the external magnetic field we define  the electric  field $\textbf{e}^{(\mathrm{c})}=\textbf{e}^{(\mathrm{1})}+\textbf{e}^{(\mathrm{3})}$ and magnetic field $\textbf{h}^{(\mathrm{c})}=-i\textbf{k}\times \textbf{a}^{(\mathrm{c})}=\textbf{h}^{(\mathrm{1})}+\textbf{h}^{(\mathrm{3})}$ associated with  $\mathbf{a}^{(\mathrm{c})}=\mathbf{a}^{(\mathrm{1})}+\mathbf{a}^{(\mathrm{3})},$ respectively.  A this point it is meaningful to analyze
\begin{eqnarray}
\boldsymbol{\mathfrak{s}}^{(\mathrm{c})}=-i\textbf{e}^{(\mathrm{c})}\times\textbf{a}^{(\mathrm{c})}\ \ \mathrm{and}\ \  \boldsymbol{\mathfrak{p}}^{(\mathrm{c})}=\textbf{e}^{(\mathrm{c})}\times\textbf{h}^{(\mathrm{c})}.\label{paserr}
\end{eqnarray}  Inserting the explicit  expression of  $\textbf{e}^{(\mathrm{c})},$  $\textbf{h}^{(\mathrm{c})}$ and $\textbf{a}^{(\mathrm{c})}$  in Eq. (\ref{paserr}) we get 
\begin{equation}
\begin{array}{c}
\displaystyle\boldsymbol{\mathfrak{s}}^{(\mathrm{c})}=\boldsymbol{\mathfrak{s}}^{(1)}+\frac{\omega\textrm{k}_\perp}{\rm k^2}\left(\textbf{n}_{\parallel} \rm
k_\perp-\textbf{n}_\perp \rm k_{\parallel}\right),\\ \\
\boldsymbol{\mathfrak{p}}^{(\mathrm{c})}=\boldsymbol{\mathfrak{p}}^{(\mathrm{1})}+\boldsymbol{\mathfrak{p}}^{(\mathrm{3})}-\omega(\textbf{k}_\perp\times\textbf{n}_\parallel).
 \end{array}
\end{equation} Note that
\begin{equation}
\lim_{\textbf{k}_\perp\to0}\boldsymbol{\mathfrak{s}}^{(\mathrm{c})}=0\ \  \mathrm{and}\ \ \lim_{\mathbf{k}_\perp\to0}\boldsymbol{\mathfrak{p}}^{(\mathrm{c})}=2\omega\textbf{k}_\parallel.\label{parpoint}
\end{equation} The resulting limits are expected for a  circularly polarized wave.

Further considering  Eq. (\ref{passsspm}) we obtain 
\begin{eqnarray}
\boldsymbol{\mathfrak{s}}^{\pm}&=&-\frac{i}{2}\left(\textbf{e}^{(c)}\pm i\textbf{h}^{(c)}\right)\times \textbf{a}^{(c)}\\&=&\frac{1}{2}\boldsymbol{\mathfrak{s}}^{(\mathrm{c})}\pm \frac{1}{2}\textbf{h}^{(\mathrm{c})}\times\textbf{a}^{(\mathrm{c})}\nonumber
\end{eqnarray} Note that
\begin{eqnarray}
\mathbf{h}^{(\mathrm{c})}\times\mathbf{a}^{(\mathrm{c})}&=&i\mathbf{k}+i
\frac{k_\parallel^2-\omega^2}{k^2}\mathbf{k}_\parallel+i\frac{k_\parallel^2}{k^2}\mathbf{k}_\perp\nonumber\\&-&i\frac{\omega^2}{k^2}\left(\mathbf{k}_\perp\times\mathbf{n}_\parallel\right).\nonumber
\end{eqnarray} The above relation  was obtained  by inserting  the  explicit form  of $\textbf{e}^{(\mathrm{c})}$, $\textbf{h}^{(\mathrm{c})}$ and $\textbf{a}^{(\mathrm{c})}.$ Therefore, the angular
momentum associated with left and right circular polarization  are
represented by
\begin{eqnarray}
\lim_{\mathbf{k}_\perp\to0}\boldsymbol{\mathfrak{s}}_{\pm}\simeq\mp \textbf{k}_\parallel.
\label{gspirmsa}
\end{eqnarray}

Regarding the normalized version of the above  limit
$(\mathbf{s}_{\pm}\equiv\mp\textbf{n}_\parallel)$  as the angular
momentum used to define the helicity,  we can write
$\lambda=\mathbf{s}_{\pm}\cdot\boldsymbol{\mathbf{p}}^{(\mathrm{c})}=\mp1,$
where  $\mathbf{p}^{(\mathrm{c})}=\mathbf{n}_\parallel$ is the
normalized version of the second  limit computed in Eq.
(\ref{parpoint}).

\subsection{Spin density  of a free electromagnetic field for perpendi\-cular propagation}

Let us  consider a free  photon  propagating perpendicular to the external magnetic  field $(\omega_{(2,3)}=\vert\textbf{k}\vert).$  In this case,  the behavior of the photon spin density   reads
\begin{equation}
\boldsymbol{\mathfrak{s}}^{(\mathrm{t})}=-i\textbf{e}^{(\mathrm{t})}\times\textbf{a}^{(\mathrm{t})}=-i\textrm{a}_{0}^{(\mathrm{t})}\textbf{h}^{(\mathrm{t})}
\end{equation} Here $\textbf{e}^{(\mathrm{t})}=\textbf{e}^{(\mathrm{2})}+\textbf{e}^{(\mathrm{3})}$ and  $\textbf{h}^{(\mathrm{t})}=\textbf{h}^{(\mathrm{2})}+\textbf{h}^{(\mathrm{3})}$ denotes  the  electric and magnetic field  associated with  $\mathrm{a}_\mu^{(\mathrm{t})}=\mathrm{a}_\mu^{(2)}+\mathrm{a}_\mu^{(3)},$ respectively. Because of  $\mathrm{a}_0^{(2)}=-\mathrm{k}_\parallel,$ $\mathrm{a}_0^{(3)}=0$  and Eq. (\ref{electric-magnetic-field}), we find 
\begin{equation}
\boldsymbol{\mathfrak{s}}^{(\mathrm{t})}= \boldsymbol{\mathfrak{s}}^{(2)}-\mathbf{n}\times\boldsymbol{\mathfrak{s}}^{(2)}
\end{equation} with $\textbf{n}=\textbf{k}/\vert\textbf{k}\vert$ being the corresponding  wave vector.

Additionally, taking  Eq. (\ref{parpoint}) into account   we  introduce the vector 
\begin{equation}
\mathbf{s}_\gamma=\left\{\begin{array}{ccc} 0&\mathrm{for}& \rm k_\perp=0\\\displaystyle
\mathbf{s}^{(2)}-\textbf{n}\times\mathbf{s}^{(2)}&\mathrm{for}& \rm k_\perp\neq0
 \end{array}\right.\label{sss1}.
\end{equation} with  $\mathbf{s}^{(2)}\equiv\boldsymbol{\mathfrak{s}}^{(2)}/\vert\boldsymbol{\mathfrak{s}}^{(2)}\vert=(\mathbf{n}_\parallel\times \mathbf{n}_\perp).$ The above expression is  total  spin  density, which itself depends  on the direction of propagation.   

Because of the fact that  $\boldsymbol{\mathbf{s}}^{(2)}\cdot\textbf{k}=0,$  the total spin density of the electromagnetic field is orthogonal to the wave vector ($\boldsymbol{\mathbf{s}}_{\gamma}\cdot\textbf{n}=0)$. We remark  that both $\mathbf{s}^{(2)}$ and $\textbf{n}\times\mathbf{s}^{(2)}$   are orthogonal to each other.  Note  that for transversal propagation  the total spin density of the electromagnetic field has a parallel component to $\textbf{B}$ given by $
\mathbf{s}_{\gamma}^{\parallel}=-\frac{\textrm{k}_\perp}{\vert\textbf{k}\vert}\textbf{n}_\parallel.$
 
\section{Photon magnetic moment}

In this section  we  explore  the possibility that  a photon might
carry a magnetic moment. We will restrict ourselves to asymptotically large magnetic  field
$\rm b=\vert\textbf{B}\vert/\rm B_c\gg 1.$   In this limit the eigenvalue of the second propagation 
mode  contains a term linearly growing with the magnetic field strength at  
low and high energy limits.   Both cases deserve a separate study because  
the $\varkappa_2-$structures  differs from one to the other energy domain.  
Under  the same conditions $\varkappa_1$ and $\varkappa_3$  cannot create a 
virtual electron-positron pair in the ground state \cite{shabad3v}.  These  show  
logarithmic dependences  on $\vert\bf B\vert$ and  their corresponding  physical  dispersion laws are 
independent of the external field (for details see Ref. \cite{shabad4}). Therefore the cases concerning to the first and third  propagation mode  are not relevant  in the current  context.  So, in this section, we will analyze 
the effects produced due to  the second  eigenmode.

\subsection{Infrared struture of $\varkappa_2$: Covariant decomposition of the photon interaction energy }

In the limit    $\rm m^2 b \gg m^2\gg\omega^2-\rm k_\parallel^2$ with $\rm m^2 b\gg \rm k_\perp^2,$ the second eigenvalues of $\Pi_{\mu\nu}$  shows  a linear function on the external field strength
\begin{equation}
\varkappa_2^{\mathrm{IR}}=-\varrho(\rm b)\rm z_1=-\frac{\alpha}{3\pi}\frac{\rm e}{\rm m^2}\rm
F_{\mu\nu}^{*}k^\mu\textrm{a}^{\nu(2)},\label{pi3}
\end{equation} with $
\varrho(\mathrm{b})=\frac{\alpha}{3\pi}\rm b$ and  $\alpha=\rm
e^2/4\pi=1/137$ being  the  fine-structure constant. Note that we have
used the decomposition:  $\rm z_1=\rm k^\mu
F_{\mu}^{*\lambda}F_{\lambda}^{*\nu}k_\nu/(2\Delta)=\rm k^\mu
F_\mu^{*\lambda}a_\lambda^{(2)\nu}/(2\Delta)^{1/2}.$  The 
expression for $\varkappa_2^{\mathrm{IR}}$ is analogous to that of the invariant interaction energy of the electron
\cite{sokolov}:
\begin{equation}
\varepsilon=\frac{\rm e}{2\rm m^2}\rm F_{\mu\nu}^*\rm
p^\mu \textrm{S}^{\nu}. 
\end{equation} Here  $\rm p^\mu$ is the electron fourth momentum and  $\rm
S^\mu=\gamma^5\left(\gamma^\mu-\frac{\Pi^\mu}{m}\right)$ is the electron 
 spin with $\rm \Pi^\mu=\rm p^\mu+\frac{i}{2} e
F^{\mu}_{\nu}x^\nu$.  In the electron rest frame, $\varepsilon$
describes the interaction energy between  the electron magnetic
moment and the external magnetic field. In the photon case the structure  of Eq. (\ref{pi3})  shows that  the electron  spin $\textrm{S}^{\mu}$ is replaced by  the photon  polarization $\rm a_\mu^{(2)}$, describing the intrinsic rotation of the photon in the plane perpendicular to the external field.  Note, however, that  the absence of a photon rest frame prevents the definition of a photon  spin, unlike the electron case. To avoid this problem and for further convenience we will  analyze the  photon propagation by investigating its  dispersion law Eq. (\ref{egg}), which  reads
\begin{equation}
\omega_2^2=\textbf{k}^2-\varrho(\rm
b)\left(1+\varrho(\mathrm{b})\right)^{-1}\rm z_2.\label{comics}
\end{equation} For magnetic field strength $10<\rm b \ll 3\pi/\alpha$ one should treat $\varrho(\rm b)$ as small. The expansion of $\omega$ up to first order in $\varrho(\rm b)$  gives the  dispersion law
\begin{equation}
\omega_2=\vert\textbf{k}\vert-\frac{1}{2} \varrho(\rm
b)\rm{z}_2/\vert\textbf{k}\vert.\label{prieq}
\end{equation}
Obviously, the first term in Eq. (\ref{prieq}) corresponds to the
light cone equation whereas the second arises due to  the dipole
moment contribution of the virtual electron-positron pair.  In  this
approximation the dispersion law does not essentially deviate  from
its vacuum shape and its interacting term grows linearly with the
external magnetic field. This fact attracts our attention  because it seems
that a magnetic moment may be  ascribed to a mode-$2$ photon. 

Hereafter we will denote the interaction energy by
\begin{equation}
\mathscr{U}=-\frac{1}{2} \frac{\varrho(\rm
b)\rm{z}_2}{\vert\textbf{k}\vert}=
\frac{\alpha}{6\pi}\frac{\rm e}{\rm m^2}\frac{(\rm z_2)^{1/2}}{\vert\bf{k}\vert}\rm{F}_{\mu\nu}\rm  k^\mu
\textrm{a}^{\nu(3)}\label{primera}
\end{equation} where the decomposition
$\textrm{z}_2=-\rm{k}^\mu \rm{F}_\mu^\nu F_\nu^\lambda
k_\lambda/(2\Delta)=(\rm z_2)^{1/2}\rm  k^\mu F_{\mu\nu}
\textrm{a}^{\nu(3)}/(2\Delta)^{1/2}$ has been used.  Note that $(\rm
z_2)^{1/2}=\rm k_\perp$  in reference frames which are at rest or moving  parallel to $\textbf{B}.$  With  this in mind, we find  by  symmetrization of Eq. (\ref{primera}) \begin{equation}
\mathscr{U}=-\frac{1}{2}\mathscr{M}_{\mu\nu}\mathrm{F}^{\mu\nu}\
\ \mathrm{with} \ \ \mathscr{M}_{\mu\nu}\equiv
-i\mathfrak{g}\frac{\rm e}{\rm 2m} f[\mathrm{k}_\perp]
\mathrm{S}_{\mu\nu}.\label{segunda}
\end{equation}Here  $
\mathfrak{g}=\frac{\alpha}{3\pi}$ is a kind
of  Land\'e factor, whereas $f[\mathrm{k}_\perp]\equiv \mathrm{k}_\perp/\mathrm{m}.$  In addition, $\mathrm{S}_{\mu\nu}\equiv
\mathfrak{F}_{\mu\nu}^{(3)}/(\vert\textbf{k}\vert)$ is  tensor with 
$\mathfrak{F}_{\mu\nu}^{(3)}=-i\rm k_\mu a_\nu^{(3)}+i\rm k_\nu
a_\mu^{(3)}$ referring to the antisymmetric electromagnetic
tensor generated  by the third propagation mode. Note that $\mathfrak{F}_{ij}^{(3)}=\epsilon_{ijk}\mathrm{h}_k.$

The spatial part of $\mathscr{M}_{\mu\nu}$  can be written as:
\begin{equation}
\mathscr{M}_{ij}=\mathfrak{g} \frac{\rm e}{\rm 2m}f[\mathrm{k}_\perp] \frac{\textrm{h}_{k}^{(3)}}{\vert\textbf{k}\vert}\mathscr{J}^{(k)}_{ij},\label{spatialpartmage}
\end{equation} 
Here $\mathscr{J}^{(i)}_{lm} =-i\epsilon_{ilm}$  is  the
$3-$dimensional  representation of the $\rm SO(3)-$generators, fulfilling 
$ \left[\mathscr{J}^{(i)},\mathscr{J}^{(j)}\right]=i\epsilon^{ijk}\mathscr{J}^{(k)}$ and $(\mathscr{J}_{lm}^{(i)})^2=\mathscr{J}_{lu}^{(i)}\mathscr{J}_{um}^{(i)}=2\delta_{lm}.$

In a Lorentz frame where $\rm F_{\mu\nu}$ is purely magnetic the
structure of Eq. (\ref{segunda})  is expanded to
\begin{equation}
\mathscr{U}=-\mathfrak{g}\frac{\rm e}{2\rm m}f[\mathrm{k}_\perp]\left[\textbf{n}_{\parallel}\sin\phi-\textbf{
n}_\perp\cos\phi\right]\cdot\textbf{B},\label{intermedia}
\end{equation} where  $0\leq\phi\leq\pi$  is the polar angle  between
the wave vector  $\bf n$ and  the external field $(\tan\phi=\rm
k_\perp/k_\parallel).$ The expression inside the brackets is a unit
vector orthogonal to  the direction of propagation.  We then write the
interaction energy as
\begin{equation}
\mathscr{U}=-\boldsymbol{m}_\gamma\cdot\textbf{B}\label{spinsas}
\end{equation} with
\begin{equation}
\boldsymbol{m}_\gamma\equiv\mathfrak{g}\frac{\rm e}{2\rm m}f[\mathrm{k}_\perp]\left(\mathbf{n} \times \mathbf{s}^{(2)}\right).\label{goddeft}
\end{equation} The struture of the interaction energy  is similar to a  potential energy of a magnetic dipole in an external magnetic field. Thus, in first approximation (with respect $\varrho(\rm b)$),  the second propagation mode in a superstrong magnetic field seems to  behave as a quasiparticle having  a magnetic moment $\boldsymbol{m}_\gamma$.  This  behavior  occurs in some scenarios of condensate matter  and  has  allowed  to introduce the concept of polariton. The latter results  from strong coupling of electromagnetic waves with a  magnetic dipole-carrying excitation. 

According to Eq. (\ref{intermedia}),  this magnetic moment is the sum of two 
orthogonal components, as dictated by the cylindrical symmetry of
our problem: the first one along the external field direction
\begin{equation}
\boldsymbol{m}^{\parallel}_{\gamma}=\mathfrak{g}\frac{\rm e}{2 \rm m}f[\mathrm{k}_\perp]
\textbf{n}_{\parallel}\sin\phi,
\end{equation}
whereas the second one is perpendicular to $\bf B$
\begin{equation}
\boldsymbol{m}^{\perp}_{\gamma}=-\mathfrak{g}\frac{\rm e}{2\rm m}f[\mathrm{k}_\perp]
\textbf{n}_\perp\cos\phi.
\end{equation}
Furthermore, both components of $\boldsymbol{m}_\gamma$ show opposite
magnetic  behavior: while the parallel one is essentially
paramagnetic  $(m^{\parallel}_\gamma>0)$, the perpendicular one is
purely diamagnetic $(m^{\perp}_\gamma<0)$. They become nonmagnetic
for propagation along ${\bf B}$, $(\rm k_\perp=0)$, in whose case
the radiation is, -like in the free Maxwell theory-, insensitive to
the magnetic field. Note that  for $\rm k_\perp\to0$ the photon anomalous magnetic
moment vanishes. This is a direct  consequence of the
gauge invariance of the eigenvalues of $\Pi_{\mu\nu}$.  
Clearly, the effective magnetic interaction is related to  $
m^\parallel_\gamma$ which is  invariant under
rotation around the external field.

\begin{figure}[!htbp]
\begin{center}
\includegraphics[width=3in]{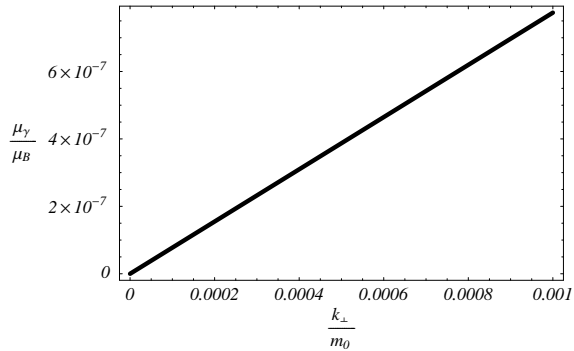}
\caption{\label{fig:mb011} Photon anomalous magnetic moment drawn
against the transverse momentum component. Here $\mu_{\rm B}=\rm
e/(2m)=9.274 \cdot 10^{-21} erg/G$ is the Bohr magneton.}
\end{center}
\end{figure}
The dependence of the photon magnetic moment on $\rm k_\perp$  is
displayed in Fig.  \ref{fig:mb011} for purely perpendicular
propagation with respect to $\textbf{B}$. Note that in this case
$\boldsymbol{m}_\gamma\cdot\textbf{n}=0.$  Within the range of
frequencies  for which $\boldsymbol{m}_\gamma$ is defined,   it is 4 orders
of magnitude smaller than the electron anomalous magnetic
moment $(\mu^{\prime}\simeq\frac{\alpha}{2\pi} \frac{\rm
e}{\rm 2m})$ \cite{Schwinger}  but it is 12 orders of magnitude larger  than the neutrino magnetic moment ($\mu_{\nu_e}=10^{-19}\mu_{\mathrm{B}}$)  \cite{Eidelman:2004wy}. However, due to  the astonishing experimental precision with which the  anomalous magnetic moment of both  the electron and muon are   measured,  there is some hope for an  experimental -probably astronomical- measurement  of the photon anomalous magnetic moment.

Note that  although the interaction energy Eq. (\ref{spinsas}) seems to be linearized on the
external field  and $\boldsymbol{m}_\gamma$ is independent of
$\vert\textbf{B}\vert$ it turns out to be a vector depending on the external field 
direction.  Therefore, if  the  photon anomalous magnetic moment is interpreted as $\boldsymbol{\mu}_\gamma=-\partial\omega/\partial \textbf{B}$, some differences with respect  to $\boldsymbol{m}_\gamma$  are expected.   However, in  Appendix \ref{hfgt} we find that 
\[ 
\boldsymbol{\mu}_\gamma\simeq\boldsymbol{m}_\gamma.
\] We would like to bring the attention that this proceedure  identifies the magnetic moment only  if the dispersion law   depends linearly  on the external magnetic field  (for  the electron case  see  Ref. \cite{Luttinger:1948zz,lipman,Suzuki:2005wra}).  Otherwise $\boldsymbol{\mu}_\gamma$ must be understood as  a sort of ``photon  magnetization'' rather than a magnetic moment. However the latter notion equips us with an  intuitive tool for qualitative analysis of the magnetization generated by a background of observable photons.

\subsection{Discussion}

We have seen   that  the external magnetic  field leaves only the 
rotational  symmetry  around  of $\bf B$ invariant. Consequently, $m^\parallel_\gamma$ is an invariant under the rotation around the axis of $\bf B$ while $m^\perp_\gamma$ does not. Note, in addition, that  $\boldsymbol{m}^\perp_{\gamma}$ does not contribute to the interaction energy. However $\boldsymbol{m}^{\perp}_\gamma$  turns out to be  relevant in another case:  the external
field does exert a torque on the magnetic dipole which tends to line 
up $\boldsymbol{m}_\gamma$  with $\textbf{B}.$ Indeed,  
\begin{equation}
\boldsymbol{\tau}_\gamma=\boldsymbol{m}_\gamma\times\textbf{B}=\boldsymbol{m}^\perp_\gamma\times\textbf{B}=\mathfrak{g}\frac{\rm e}{\rm 2m}f[\rm k_\perp]\vert\mathbf{B}\vert\cos\phi\textbf{s}^{(2)}.\label{torque}
\end{equation} Because of the fact that  the torque is collinear to $\textbf{s}^{(2)}$, the transversal components of  total angular momentum  in that direction is not conserved. This is a manifestation of the  reduction  of the rotation symmetry and, consequently  Lorentz symmetry. Therefore $\boldsymbol{m}_\gamma$ is a signal of a break-down of the Lorentz symmetry. This fact corroborates and extends the result presented in Ref. \cite{Belich:2008vz} which   claims that a magnetic dipole moment of truly-elementary massive neutral particles is a  signal of Lorentz  symmetry violation. Note, in addition, that   $\boldsymbol{\tau}_\gamma$ is orthogonal to
$\textbf{n}\times \textbf{s}^{(2)}$  which  might causes a precession of
$\boldsymbol{m}_{\gamma}$ around $\textbf{B}$ in  analogy to the
Larmor precession for  electrons. 

The  Lorentz symmetry breaking  implies that  the photon helicity is no longer a conserved quantity due to rotations transversal  to $\textbf{B}$ (see Subsect. \ref{spingenerallrofjdffk}).  As a consequence,  alternative conserved  elements   are needed to  identify   a  photon interacting with $\textbf{B}$.  Once  the magnetic moment is a coefficient of the linear contribution in the magnetic field it may become a characteristic  quantity since one may attribute it an ``spin.''  However this is only  a possible point of view and  only further studies can tell  how far  the interpretation can be stretched. A realistic treatment of this problem  requires a group theoretical analysis  including $\boldsymbol{m}_\gamma$ similar  to  those developed by Wigner for a free photon \cite{Wigner:1939cj}. The latter is beyond the  scope of this manuscript, but a study in this direction  is in progress. 

Let us consider now  a system consisting of $\mathscr{N}$ independent photons 
described by the dispersion equation 
\begin{equation}
\omega=\vert\textbf{k}\vert-\boldsymbol{m}_{\gamma}\cdot\bf
B\label{principal12323}
\end{equation}where we have considered  Eq. (\ref{prieq}),  Eq. (\ref{primera}) and  Eq. (\ref{spinsas}).
In addition we will suppose that they propagate  purely transversal  to the external
magnetic field. According to these assumptions, the total
interaction energy of the system is 
\begin{eqnarray}
\mathscr{U}_{\mathrm{total}}=-\sum_{i=i}^{\mathscr{N}}m_\gamma^{\parallel(i)}\vert\mathbf{B}\vert=-\mathscr{N}\langle m_\gamma^{\parallel}\rangle\vert\mathbf{B}\vert
\end{eqnarray} where  $m_\gamma^{(i)}$ denotes the magnetic  moment of each photon. Here  the average of the photon  magnetic moment is   $\langle m_\gamma^{\parallel}\rangle=\sum_{i=i}^{\mathscr{N}}m_\gamma^{\parallel(i)}/\mathscr{N}\sim\langle\mathrm{k}_\perp\rangle.$  For a uniform  photon distribution in the transversal
plane to $\mathbf{B},$ one  expects that
$\langle\mathrm{k}_\perp\rangle=0$   which implies that  $\mathscr{U}_{\mathrm{total}}=0$. As a consequence,   the system does not carry a magnetization $\boldsymbol{\mathfrak{M}}_\gamma=-\partial\mathscr{U}_{\mathrm{total}}/\partial \mathbf{B}.$ If the system  is  considered as a monochromatic beam $\langle\mathrm{k}_\perp\rangle\neq0,$ $\mathscr{U}_{\mathrm{total}}\neq0$  and  
\begin{eqnarray}
\vert\boldsymbol{\mathfrak{M}}_\gamma\vert&=&-\partial \mathscr{U}_{\mathrm{total}}/\partial \vert\mathbf{B}\vert=\mathscr{N}\langle m_\gamma^{\parallel}\rangle\\&=&\mathfrak{g}\frac{\rm e}{\rm 2 m}\mathscr{N}\langle f[\mathrm{k}_\perp]\rangle, \ \ \ \ \langle f[\mathrm{k}_\perp]\rangle=\langle\mathrm{k}_\perp\rangle/\mathrm{m}\nonumber
\end{eqnarray} In consequence  the beam carries a nonzero magnetization
 which alters the external field $\bf  B$:
\[\bf H=\textbf{B}+4\pi(\boldsymbol{\mathfrak{M}}_\gamma+\boldsymbol{M}_{\mathrm{vac}}).\] Here  $\boldsymbol{M}_\mathrm{vac}>0$ is the vacuum magnetization (for details see Sect. \ref{sec:magnetvacuum}).  Within the  magnetic field interval for which $\boldsymbol{m}_\gamma$ is defined,   $\vert\boldsymbol{M}_\mathrm{vac}\vert\sim10^{10}-10^{12} \rm erg/(cm^3 G).$ Now, in the frequency range of  x-rays $\rm (\langle k_\perp\rangle\geq 150\ \ eV)$   the averaged  magnetic moment $\langle m_\gamma\rangle\sim10^{-27}\rm erg/G.$ In order to  produce a photon magnetization of the order of  $\sim \vert\boldsymbol{M}_\mathrm{vac}\vert,$   a photon density  of order $\rho_\gamma=\mathscr{N}/\mathrm{V}\sim10^{36}-10^{38}\ \ \rm cm^{-3}$ would be necessary. Moreover, for $\rho_\gamma\gtrsim10^{42}\rm cm^{-3}$ the magnetization carried by the beam is $\vert\boldsymbol{\mathfrak{M}}_\gamma\vert\sim10^{15}\rm erg/(cm^3G)$ and thus larger than $\vert\boldsymbol{M}_\mathrm{vac}\vert$ even more, it is 1 order of magnitude larger than the external magnetic field $\vert\textbf{B}\vert\sim10^{14}\rm G/cm^3$.
 
Finally, we want to conclude this subsection pointing out that Eq. (\ref{principal12323})  allows to determine the  the corresponding vacuum  refraction index  in terms of the photon magnetic moment:
\begin{equation}
\rm n=\frac{\vert \textbf{k}\vert}{\omega}\simeq1+\frac{\boldsymbol{m}_\gamma\cdot\bf B}{\vert \textbf{k}\vert}.\label{refraccionindex}
\end{equation}  The latter expression  depends on the direction of  the photon momentum  and reaches its maximum for a purely transversal propagation ($\rm k_\parallel=0$) in which case 
$\rm n_{\mathrm{max}}\simeq 1+\frac{\alpha}{3\pi}\frac{\rm e}{2\rm m^2}\vert\mathbf{B}\vert.$ For $\rm b\sim 100,$  $\rm n_{\mathrm{max}}\approx1.038$  which exceeds the values of typical gases at  atmospheric pressure in absence of $\textbf{B}$ \cite{shabad4}. We want to remark that  Eq. (\ref{refraccionindex}) might play an important role in the evaluation of the magnetic effect on  gravitational lenses in the vicinity  of highly magnetized  stellar objects \cite{Denisov:2001gv,Denisov:2005si}.

\subsection{Ultraviolet domain\label{ultravioletmu}}

Next we consider  the  high-energy regime. In the limit $\rm
m^2b\gg\omega^2-\textrm{k}_\parallel^2\to\infty$ and   $\rm m^2b\gg
\textrm{k}_\perp^2,$  the second eigenvalue approaches
\begin{equation}
-\varkappa_2^{\mathrm{UV}}=\mathfrak{m}_2^2=\frac{2\alpha}{\pi}\rm e\vert\textbf{B}\vert.
\end{equation} Here  $\mathfrak{m}_2$ is the photon effective mass \cite{Osipov}
corresponding to the topological one  in the  $2-$dimensional
Schwinger model \cite{Schwinger2}.  Its  derivation is closely
related to the chiral limit in which the axial current is not
conserved (for more details we refer the reader to Ref.
\cite{salim}). As a  consequence, an ultraviolet  photon seems to
behave like a neutral massive vector boson whose movement is
quasiconfined in $1+1$ dimensions
\begin{equation}
\omega\approx\left(\rm k_\parallel^2+\mathfrak{m}_2^2\right)^{1/2}.\label{quasi1}
\end{equation}
In contrast to the previous case $\varkappa_2^{\mathrm{UV}}$ does
not depend  explicitly on the polarization mode. This fact is also
manifest  in the dispersion law which cannot be linearized in the external magnetic field  and therefore 
a photon magnetic moment is not expected.  However, this kind of photon carries a magnetization density  given by 
\begin{equation}
\boldsymbol{\mu}_{\gamma}=-\left.\frac{\partial \omega}{\partial
\mathbf{
B}}\right\vert_{\mathbf{B}_\perp\to0}=-\frac{\alpha}{\pi}\frac{\rm
e}{\omega}\textbf{n}_\parallel.\label{mquasi}
\end{equation}

Note that $\boldsymbol{\mu}_{\gamma}<0$ 
behaves  diamagnetically  and depends on
the external field strength.  In particular for
$\rm k_\parallel^2\ll \mathfrak{m}_2^2$
\begin{equation}
\boldsymbol{\mu}_{\gamma}\simeq-\frac{\alpha}{\pi}\frac{\rm
e}{\mathfrak{m}_2}\textbf{n}_\parallel=-\left(\frac{2\alpha}{\pi\mathrm{b}}\right)^{1/2}\mu_\mathrm{B}\textbf{n}_\parallel,
\end{equation}
which tends to vanish for $\vert\textbf{B}\vert\to\infty.$ 

For  $\mathrm{b}\sim10^3$ corresponding to magnetic field $\vert\mathbf{B}\vert\sim 10^{16}\mathrm{G}$ the photon  magnetization  of an ultraviolet radiation   reaches values  of the order of $\mu_\gamma\sim-\mu_\mathrm{B}.$ For  the same magnetic field strength, an infrared photon propagating  perpendicular to $\bf B$ has $\mu_\gamma^{\parallel}\simeq\mathfrak{g}\frac{\rm e}{\rm 2 m}\varrho(\mathrm{b})^{-5/2}\sim10^{-12}\mu_\mathrm{B}$ (see Appendix \ref{hfgt}). As a consequence, the magnetic response of a photon background compoused  by an equal number of infrared an ultraviolet  radiations  will be dominated  by the  ultraviolet contribution, which  for mode -$2$  is diamagnetic. The latter behavior is also manifest in the contribution of this eigenmode to the vacuum magnetization density (see Sect. \ref{sec:magnetvacuum}).

\section{Two loop term  of the  Euler-Heisenberg Lagrangian}

Virtual photons  can interact  with the external magnetic field by
means of the vacuum polarization tensor. As a consequence they
might be a source of magnetization  to the whole vacuum. In what
follows we  compute  this contribution for  very large magnetic
fields ($\mathrm{b}\gg 1$), which might exist  in stellar objects like
neutron stars.

\subsection{The unrenornamalized contribution due to  the vacuum polarization eigenvalues}

We start our analysis with  the  Euler-Heisenberg  Lagrangian
\begin{equation}
\mathfrak{L}_{\mathrm{EH}}=\mathfrak{L}_{\mathrm{R}}^{(0)}+\mathfrak{L}_{\mathrm{R}}^{(1)}+\ldots\label{therdp}
\end{equation} with
\begin{equation}
\mathfrak{L}_{\mathrm{R}}^{(0)}=-\frac{1}{2}\mathrm{B}^2\ \ \mathrm{and}\ \ \mathrm{B}=\mathrm{B}_0\mathcal{Z}_{3(1loop)}^{-1/2}.
\end{equation} Here $\mathfrak{L}_{\mathrm{R}}^{(0)}$ is the free  renormalized Maxwell Lagrangian in a 
Lorentz  frame where the electric field vanishes, $\textbf{E}=0$.
Hereafter $\rm B_0$, $\rm m_0$ and $\rm e_0$  will be referred as
the ``bare magnetic  field strength'', ``bare electron mass'' and
``bare charge'',  respectively.  Without the index
$0$ these quantities   must be understood as renormalized. On the other hand, 
$\mathfrak{L}_{\mathrm{R}}^{(1)}$ is  the  regularized one-loop  
contribution  due to the  virtual electron-positron pairs created and
annihilated spontaneously in vacuum and interacting with the
external field \cite{euler}:
\begin{equation}
\mathfrak{L}_{\mathrm{R}}^{(1)}= -\frac{1}{8\pi^2}\int_0^\infty\frac{\rm d \tau}{\rm \tau^3}\rm e^{-\rm m_0^2
\tau } \left(\rm s \coth(\rm s)-1-\frac{\rm s^2}{3}\right)\label{1looprenor}
\end{equation} with $\rm s=eB \tau$ and $\rm e=e_0\mathcal{Z}_3^{1/2}.$ In this
context, the one-loop renormalization constant is given by
\begin{equation}
\mathcal{Z}_{3(1loop)}^{-1}=\lim_{\tau_0\to0}\left\{1+\frac{\alpha}{3\pi}\ln\left(\frac{1}{\gamma \rm m_0^2\tau_0}\right)\right\},
\end{equation} where $\ln(\gamma)=0.577\ldots$ is the Euler constant.

\begin{figure}[!htbp]
\begin{center}
\includegraphics[width=2.5in]{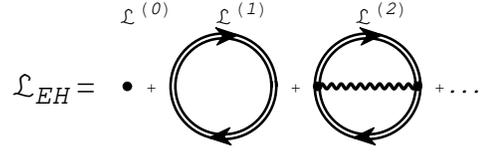}
\caption{\label{fig:mb000} Two-loop expansion  of the
Euler-Heisenberg Lagrangian. The solid lines represent the electron-positron Green's functions, whereas the wavy line refers to the
photon. Here $\mathfrak{L}^{(1)}$ is represented by the one-loop graph
which gives the contribution of  the virtual free electron-positron
pairs created and annihilated spontaneously in vacuum and
interacting with the external field. The radiative corrections
(involved in  $\mathfrak{L}^{(2)}$) emerge from  the two-loop graph due to
exchange of  virtual photons.}
\end{center}
\end{figure}

The  contribution  due to  virtual photons interacting with the external
field  by means of the vacuum polarization tensor is expressed as
\begin{equation}
\mathfrak{L}^{(2)}=\frac{i}{2}\int\frac{\rm
d^4k}{(2\pi)^4}\Pi_{\mu\nu}(\rm k)\mathfrak{D}^{\mu\nu}(\rm k).
\end{equation}
Substitution of Eq. (\ref{gstrpi}) and  Eq. (\ref{photonpro}) into the latter  expression gives
\begin{equation}
\mathfrak{L}^{(2)}=\frac{i}{2}\sum_{i=1}^3\int\frac{\rm d^4k}{(2\pi)^4}\frac{\varkappa_i}{(\rm k^2-\varkappa_i)}.
\end{equation} Manifestly, this expression  shows that each photon propagation
mode contributes independently. The quantity $\Pi_{\mu\nu}(\rm k)\mathfrak{D}^{\mu\nu}(\rm k)=\sum_i\varkappa_i(\rm k^2-\varkappa_i)^{-1}$ represents the intercation energy between the full photon propagator with the vacuum polarization tensor. In order to obtain  the leading term in an expansion    in  powers of $\mathrm e^2$, i. e. the   two-loop contribution as  shown in Fig. (\ref{fig:mb000}), we just neglect  $\varkappa_i$ in the denominator of the photon propagator. Hereafter we consider this  approximation, in which  case
\begin{equation}
\mathfrak{L}^{(2)}=\sum_{i=1}^3\mathfrak{L}_i^{(2)} \ \ \mathrm{with}\ \ \mathfrak{L}_i^{(2)}\equiv \frac{i}{2}\int\frac{\rm d^4k}{(2\pi)^4}\frac{\hat{\varkappa}_i}{\rm  k^{2}}.\label{general2lterm}
\end{equation} Note that $\hat{\varkappa}_i$ denotes the unrenormalized eigenvalue.
Its expression is given by Eqs. (\ref{poleig1}-\ref{poleig4}) but
considering  only the first line of Eq. (\ref{poleig2}) where  all
physical parameters are unrenormalized.

\begin{figure}[!htbp]
\begin{center}
\includegraphics[width=2.5in]{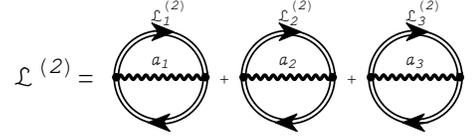}
\caption{\label{fig:mb0fgh00} Diagrammatic decomposition of
$\mathfrak{L}^{(2)}$  in terms of the vacuum polarization
eigenmodes.}
\end{center}
\end{figure}

Since the analytical properties of any  $\varkappa_i$ differs
from the others, each  contribution
$\mathfrak{L}_i^{(2)}$ will be different and therefore,  the original
$2-$loop graph can be  decomposed into three diagrams (see Fig.
\ref{fig:mb0fgh00}). Since we are interested in  the magnetic
properties generated by each photon propagation mode, each term  
$\mathfrak{L}_i^{(2)}$ will be  studied separately.

The symmetry of our problem suggests to perform the ${\rm k}$
integration with  two sets of polar coordinates corresponding to  the two planes
$(\rm k_x, k_y)$ and $(\rm k_\parallel,k_4)$, with $\rm
k_4=i\omega$. However, it is  convenient to use  the integration
variables $\rm z_2=\rm k_\perp^2$ and $\rm z_1\equiv \rm
k_\parallel^2+k_4^2>0, $ such  that
\begin{equation}
\mathfrak{L}_i^{(2)}=\frac{1}{32\pi^2}\int_0^\infty\rm  d \ell \rm d z_2 \rm d z_1 \hat{\varkappa}_i(\rm
z_2,z_1)\rm e^{-\ell(\rm z_2+z_1)}.\label{ASSA}
\end{equation}
Substitution of $\hat{\varkappa}_i$ in Eq. (\ref{ASSA})  yields:
\begin{equation}
\mathfrak{L}_i^{(2)}=-\frac{\alpha_0}{32\pi^3}\int_0^\infty\rm d\tau\exp(-\rm m_0^2\tau)\int_{0}^1 \rm d\eta\mathcal{Q}_i(\rm s,\eta)\label{remdiveprc}
\end{equation} with $\rm s=e_0B_0\tau$ and
\begin{eqnarray}
\mathcal{Q}_1(\rm s,\eta)&=&\frac{\sigma_1}{\sinh\rm s}\left\{\mathtt{V}(\rm s,\eta)+\mathtt{W}(\rm s,\eta)\right\},\\
\mathcal{Q}_2(\rm s,\eta)&=&\frac{\sigma_2}{\sinh\rm s}\mathtt{V}(\rm s,\eta)+\frac{\sigma_1}{\sinh\rm s}\mathtt{W}(\rm s,\eta),\\\mathcal{Q}_3(\rm s,\eta)&=&\frac{\sigma_1}{\sinh\rm s}\mathtt{V}(\rm s,\eta)+\frac{\sigma_3}{\sinh\rm s}\mathtt{W}(\rm s,\eta).
\end{eqnarray}The functions $\mathtt{V}(\rm s,\eta)$ and $\mathtt{W}(\rm s,\eta)$ are
given by the following integral representations
\begin{eqnarray}
\mathtt{V}(\rm s,\eta)&=&\int_0^\infty\rm  d \ell \rm d z_2 \rm d z_1 z_1\rm e^{-\rm z_2\left(\frac{\rm M}{\mathrm{eB}}+\ell\right)-\rm z_1\left(\frac{\rm N}{\mathrm{B}}+\ell\right)},\\
\mathtt{W}(\rm s,\eta)&=&\int_0^\infty\rm  d \ell \rm d z_2 \rm d z_1 z_2\rm e^{-\rm z_2\left(\frac{\rm M}{eB}+\ell\right)-\rm z_1\left(\frac{\rm N}{eB}+\ell\right)}.
\end{eqnarray}  These  integrals can be explicitly calculated:
\begin{eqnarray}
\mathtt{V}(\rm s,\eta)&=&\frac{2\rm e_0^3\rm B_0^3}{(\rm M-N)N}-\frac{2\rm e_0^3\rm B_0^3}{(\rm M-N)^2}\ln\left(\frac{\rm M}{\rm N}\right),\\\mathtt{W}(\rm s,\eta)&=&\frac{2\rm e_0^3\rm B_0^3}{(\rm N-M)M}+\frac{2\rm e_0^3\rm B_0^3}{(\rm M-N)^2}\ln\left(\frac{\rm M}{\rm N}\right),
\end{eqnarray}such that 
\begin{equation}
\mathtt{V}(\rm s,\eta)+\mathtt{W}(\rm s,\eta)=2\rm e_0^3\rm B_0^3\rm M^{-1}(\rm s,\eta)N^{-1}(\rm s,\eta). 
\end{equation}

\subsection{Renormalized contributions due to the photon polarization modes}
While the $\tau-$integral  in Eq. (\ref{remdiveprc}) does not
diverge at $\tau \to \infty,$ the  integrand is singular for  
 $\tau\to0.$ In order to  regularize $\mathfrak{L}_i^{(2)}$ we introduce a finite lower limit
$\tau_0>0$ for the proper time integral of the electron propagators
contained in the polarization tensor. As a consequence,  we can
write
\begin{equation}
\mathfrak{L}_i^{(2)}=-\frac{\alpha_0}{32\pi^3}\int_{2\tau_0}^\infty\rm d\tau\rm e^{-\rm m_0^2\tau}\int_{0}^{\eta_0} \rm d\eta\left\{\mathcal{Q}_i(\rm s,\eta)-\mathcal{Q}_{i0}(\tau,\eta)\right\}\label{remdiveprc1}
\end{equation}with  $\eta_0=1-2\tau_0/\tau$ and 
\begin{equation}
\mathcal{Q}_{i0}(\rm s,\eta)=\frac{8(\rm e_0B_0)^3}{(1-\eta^2)\rm s^3}.
\end{equation}The substraction of this term guarantees  that
$\mathfrak{L}_i^{(2)}=0$ for vanishing  magnetic fields. 
Obviously,  Eq. (\ref{remdiveprc1}) differs from  the original 
two-loop contributions in a term which is magnetic field independent
and therefore a constant.

We proceed by adding  and subtracting  the functions $\mathcal{Q}_{i2}(\rm s,\eta)$ in the
integrand  Eq. (\ref{remdiveprc1}), such that
\begin{eqnarray}\label{2remdiveprl}
\mathfrak{L}_{i}^{(2)}&=&-\frac{\alpha_0}{32\pi^3}\int_{2\tau_0}^\infty\rm d\tau\rm e^{-\rm m_0^2\tau}\int_{0}^{\eta_0} d\eta\mathcal{Q}_{i2}(\rm s,\eta)\\&-&\frac{\alpha_0}{32\pi^3}\int_{2\tau_0}^\infty\rm d\tau\rm e^{-\rm m_0^2\tau}\int_{0}^{\eta_0} d\eta\left\{\mathcal{Q}_i(\rm s,\eta)-\mathtt{Q}_{i}(\rm s,\eta)\right\}\nonumber
\end{eqnarray}with
\begin{equation}
\begin{array}{c}
\displaystyle\mathcal{Q}_{12}(\rm s,\eta)=-\frac{2\rm (e_0B_0)^3}{3\rm s},\ \ \mathcal{Q}_{22}(\rm s,\eta)=\frac{4\rm  (e_0B_0)^3}{3\rm s} \frac{1}{1-\eta^2},\\ \\ \displaystyle \mathcal{Q}_{32}(\rm s,\eta)=-\frac{\rm  (e_0B_0)^3}{3\rm s}- \frac{4\rm  (e_0B_0)^3}{3\rm s}\frac{1}{1-\eta^2}.
\end{array}\nonumber
\end{equation} Note that  the function  $\mathtt{Q}_{i}(\rm s,\eta)\equiv\mathcal{Q}_{i0}(\rm s,\eta)+\mathcal{Q}_{i2}(\rm s,\eta)$ is the expansion of $\mathcal{Q}_i(\rm s,\eta)$ up to quadratic terms in the external  field.  Obviously, for $\tau_0\to0$ the first term in Eq. (\ref{2remdiveprl}) is logarithmically  divergent $(\sim\ln\tau_0^{-1}).$ This divergence  will be ``reabsorbed''  in the course of charge renormalization.

To regularize the remaining integration over  $\eta$  we  express
Eq. (\ref{2remdiveprl}) in the following way
\begin{eqnarray}\label{3remdive}
\mathfrak{L}_{i}^{(2)}&=&-\frac{\alpha_0}{32\pi^3}\int_{2\tau_0}^\infty\rm d\tau\rm e^{-\rm m_0^2\tau}\int_{0}^{\eta_0} d\eta\mathcal{Q}_{i2}(\rm s,\eta)\\&-&\frac{\alpha_0}{32\pi^3}\int_{2\tau_0}^\infty\rm d\tau\rm e^{-\rm m_0^2\tau}\int_{0}^{\eta_0} d\eta\mathfrak{H}_{i}(\rm s,\eta)\nonumber\\&-&\frac{\alpha_0}{32\pi^3}\int_{0}^\infty\rm d\tau\rm e^{-\rm m_0^2\tau}\int_{0}^{1} d\eta\left\{\mathcal{Q}_i-\mathtt{Q}_{i}-\mathfrak{H}_i\right\}\nonumber
\end{eqnarray}  where  $\mathfrak{H}_i(\rm s,\eta)$ is a 
function  determined by the singular  part of the  Laurent series
of $\mathcal{Q}_i(\rm s,\eta)-\mathtt{Q}_{i}(\rm s,\eta)$ at
$\eta=1$. As a consequence,  the integrations in
the third line of Eq. (\ref{3remdive}) converges for $\tau_0\to0$. In particular,
\begin{eqnarray}
\mathfrak{H}_{1}(\rm s,\eta)&=&\frac{4\rm (e_0B_0)^3}{\rm s^3(1-\eta^2)}\left(\rm \rm s \coth(\rm  s)+\frac{\rm  s^2}{\sinh^2(\rm s)} -2\right),\\
\mathfrak{H}_{2}(\rm s,\eta)&=&\frac{2(\rm e_0B_0)^3}{\rm s^3(1-\eta^2)}\left(3\rm s \coth(\rm  s)+\frac{\rm s^2}{\sinh^2(\rm s)}-\frac{2 \rm s^2}{3}-4\right)\nonumber,\\
\mathfrak{H}_{3}(\rm s,\eta)&=&\frac{2(\rm e_0B_0)^3}{\rm s^3(1-\eta^2)}\left(\rm s \coth(\rm s)+\frac{3\rm s^2}{\sinh^2(\rm  s)}+\frac{2 \rm s^2}{3}-4\right)\nonumber
\end{eqnarray}

We consider the integration  over $\eta$ in the first and second
line of  Eq. (\ref{3remdive}) to express (after some algebraical
manipulations)
\begin{eqnarray}
\mathfrak{L}_{1}^{(2)}&=&-\frac{\alpha_0^2}{6\pi^2}\mathfrak{L}^{(0)}\int_{2\tau_0}^\infty\rm \frac{d\tau}{\tau}\rm e^{-\rm m_0^2\tau}+\frac{1}{3}\delta\rm m^2\frac{\partial \mathfrak{L}_\mathrm{R}^{(1)}}{\partial \rm m_0^2}+\mathfrak{L}_{1\mathrm{R}}^{(2)},\label{l1dt}\\
\mathfrak{L}_{2}^{(2)}&=&-\frac{\alpha_0^2}{6\pi^2}\mathfrak{L}^{(0)}\left[\ln\left(\gamma\rm m_0^2\tau_0\right)\int_{2\tau_0}^\infty\rm \frac{d\tau}{\tau}\rm e^{-\rm m_0^2\tau}\right.\nonumber\\&-&\left.\int_{2\tau_0}^\infty\rm \frac{d\tau}{\tau}\rm e^{-\rm m_0^2\tau}\ln\left(\gamma\rm m_0^2\tau\right)\right]-\frac{\alpha_0}{2\pi}\ln\left(\gamma\rm m_0^2\tau_0\right)\mathfrak{L}_{\mathrm{R}}^{(1)}\nonumber\\&+&\frac{1}{6}\delta\rm m^2\frac{\partial \mathfrak{L}_\mathrm{R}^{(1)}}{\partial \rm m_0^2}+\mathfrak{L}_{2\mathrm{R}}^{(2)},\label{l2dt}\\
\mathfrak{L}_{3}^{(2)}&=&-\frac{\alpha_0^2}{12\pi^2}\mathfrak{L}^{(0)}\int_{2\tau_0}^\infty\rm \frac{d\tau}{\tau}\rm e^{-\mathrm{m}_0^2\tau}+\frac{\alpha_0^2}{6\pi^2}\mathfrak{L}^{(0)}\left[\ln\left(\gamma\mathrm{m}_0^2\tau_0\right)\right.\nonumber\\&\times&\left.\int_{2\tau_0}^\infty\rm \frac{d\tau}{\tau}\mathrm {e}^{-\mathrm{m}_0^2\tau}-\int_{2\tau_0}^\infty \frac{\mathrm{d}\tau}{\tau}\mathrm{e}^{-\mathrm{m}_0^2\tau}\ln\left(\gamma\mathrm{m}_0^2\tau\right)\right]\nonumber\\&+&\frac{\alpha_0}{2\pi}\ln\left(\gamma\mathrm{m}_0^2\tau_0\right)\mathfrak{L}_{\mathrm{R}}^{(1)}+\frac{1}{2}\delta\mathrm{m}^2\frac{\partial \mathfrak{L}_\mathrm{R}^{(1)}}{\partial \rm m_0^2}+\mathfrak{L}_{3\mathrm{R}}^{(2)}\label{l3dt}
\end{eqnarray} with $\mathfrak{L}^{(0)}=-1/2 \rm B_0^2.$ Here
\begin{equation}
\delta\mathrm{m}^2=\frac{3\alpha_0\mathrm{m}_0^2}{2\pi}\left[\ln\left(\frac{1}{\gamma\mathrm{m}_0^2 \tau_0}\right)+\frac{5}{6}\right]\label{massdes}
\end{equation} is  the correction to the square  of the ``bare'' electron mass
$(\rm m^2=\rm m_0^2+\delta m^2)$ \cite{Schwinger,ritus,ritus1,dittrich1}, whereas the renormalized two loop term $\mathfrak{L}_{i\mathrm{R}}^{(2)}$ corresponding to  each eigenmode reads
\begin{eqnarray}
\mathfrak{L}_{1\mathrm{R}}^{(2)}&=&-\frac{5\alpha_0\rm m_0^2}{12\pi}\frac{\partial \mathfrak{L}_\mathrm{R}^{(1)}}{\partial \rm m_0^2}-\frac{\alpha_0}{16\pi^3}\int_{0}^\infty\rm \frac{d\tau}{\tau^3}\rm e^{-\rm m_0^2\tau}\ln\left(\gamma\rm m_0^2\tau\right)\nonumber\\ &\times&\left[\rm s\coth(\rm s)+\frac{\rm s^2}{\sinh^2(\rm s)}-2\right]-\frac{\alpha_0}{32\pi^3}\int_{0}^\infty\rm d\tau\rm e^{-\rm m_0^2\tau}\nonumber\\&\times&\int_{0}^{1} d\eta\left\{\mathcal{Q}_1(\rm s,\eta)-\mathtt{Q}_{1}(\rm s,\eta)-\mathfrak{H}_1(\rm s,\eta)\right\},\\
\mathfrak{L}_{2\mathrm{R}}^{(2)}&=&-\frac{5\alpha_0\rm m_0^2}{24\pi}\frac{\partial \mathfrak{L}_\mathrm{R}^{(1)}}{\partial \rm m_0^2} -\frac{\alpha_0}{32\pi^3}\int_{0}^\infty\rm \frac{d\tau}{\tau^3}\rm e^{-\rm m_0^2\tau}\ln\left(\gamma\rm m_0^2\tau\right)\nonumber\\&\times&\left[3\rm s\coth(\rm s)+\frac{\rm s^2}{\sinh^2(\rm s)}-4-\frac{2\rm s^2}{3}\right]-\frac{\alpha_0}{32\pi^3}\int_{0}^\infty\rm d\tau\nonumber\\&\times&\rm e^{-\rm m_0^2\tau}\int_{0}^{1} d\eta\left\{\mathcal{Q}_2(\rm s,\eta)-\mathtt{Q}_{2}(\rm s,\eta)-\mathfrak{H}_2(\rm s,\eta)\right\},\\
\mathfrak{L}_{3\mathrm{R}}^{(2)}&=&-\frac{15\alpha_0\rm m_0^2}{24\pi}\frac{\partial \mathfrak{L}_\mathrm{R}^{(1)}}{\partial \rm m_0^2}-\frac{\alpha_0}{32\pi^3}\int_{0}^\infty\rm \frac{d\tau}{\tau^3}\rm e^{-\rm m_0^2\tau}\ln\left(\gamma\rm m_0^2\tau\right)\nonumber\\&\times&\left[\rm s\coth(\rm s)+\frac{3\rm s^2}{\sinh^2(\rm s)}-4+\frac{2\rm s^2}{3}\right]-\frac{\alpha_0}{32\pi^3}\int_{0}^\infty\rm d\tau\nonumber\\&\times&\rm e^{-\rm m_0^2\tau}\int_{0}^{1} d\eta\left\{\mathcal{Q}_3(\rm s,\eta)-\mathtt{Q}_{3}(\rm s,\eta)-\mathfrak{H}_3(\rm s,\eta)\right\}.
\end{eqnarray}

Inserting Eqs. (\ref{l1dt}-\ref{l3dt}) into Eq.
(\ref{general2lterm}),  the two-loop correction to the
Euler-Heisenberg Lagrangian is given by 
\begin{equation}
\mathfrak{L}^{(2)}=-\frac{\alpha_0^2}{4\pi^2}\mathfrak{L}^{(0)}\int_{2\tau_0}^\infty\rm \frac{d\tau}{\tau}\rm e^{-\rm m_0^2\tau}+\delta\rm m^2\frac{\partial \mathfrak{L}_\mathrm{R}^{(1)}}{\partial \rm m_0^2}+\mathfrak{L}_{\mathrm{R}}^{(2)}\label{apunto}
\end{equation} with  $\mathfrak{L}_{\mathrm{R}}^{(2)}=\sum_{i=1}^3\mathfrak{L}_{i\mathrm{R}}^{(2)}.$

Considering Eq. (\ref{therdp}) and   Eq. (\ref{apunto}),   we obtain
\begin{eqnarray}
\mathfrak{L}_{\mathrm{EH}}=\mathfrak{L}^{(0)}\mathcal{Z}_{3(2loop)}^{-1}+\mathfrak{L}_{\mathrm{R}}^{(1)}+\delta\rm m^2\frac{\partial \mathfrak{L}_\mathrm{R}^{(1)}}{\partial \rm m_0^2}+\mathfrak{L}_{\mathrm{R}}^{(2)}
\end{eqnarray}where
\begin{eqnarray}
\mathcal{Z}_{3(2loop)}^{-1}&=&\mathcal{Z}_{3(1loop)}^{-1}-\lim_{\tau_0\to0}\frac{\alpha_0^2}{4\pi^2}\int_{2\tau_0}^\infty\rm \frac{d\tau}{\tau}\rm e^{-\rm m_0^2\tau}.
\end{eqnarray} For $\tau_0\to0$ the integral
$\int_{2\tau_0}^\infty\rm \frac{d\tau}{\tau}\rm e^{-\rm
m_0^2\tau}=\ln(2\gamma\rm m_0^2\tau_0)^{-1}$  and therefore
\begin{eqnarray}  
\left.\mathcal{Z}_{3(2loop)}^{-1}\right\vert_{\tau_0\to0}=1+\frac{\alpha_0}{3\pi}\ln\left(\frac{1}{\gamma\rm m_0^2\tau_0}\right)-\frac{\alpha_0^2}{4\pi^2}\ln\left(\frac{1}{2\gamma\rm m_0^2\tau_0}\right)\nonumber
\end{eqnarray}

Now, we are able to identify
\begin{equation}
\mathfrak{L}_{\mathrm{R}}^{(1)}(\rm m^2)=\mathfrak{L}_{\mathrm{R}}^{(1)}(\rm m_0^2)+\delta\rm m^2\left.\frac{\partial \mathfrak{L}_\mathrm{R}^{(1)}}{\partial \rm m_0^2}\right\vert_{\delta\rm m^2=0}
\end{equation} where  $\rm m$ is the renormalized electron mass.
The renormalized charge and field strength are introduced
by means of the  relations $\rm e=e_0\mathcal{Z}_{3(2loop)}^{1/2}$  and
$\rm B=B_0\mathcal{Z}_{3(2loop)}^{-1/2}.$ Under this condition
$\mathfrak{L}^{(0)}_{\mathrm{R}}=\mathfrak{L}^{(0)}\mathcal{Z}_{3(2loop)}^{-1}.$
Note that  $\rm e B=e_0B_0.$ So, the variable $\rm s$ is an
invariant under the renormalization. Clearly, $\rm m_0$ must be
replaced by $\rm m$ wherever it appears as well as
$\alpha_0\to\alpha$. Keeping this in mind,
\begin{equation}
\mathfrak{L}_{\mathrm{EH}}=\mathfrak{L}_{\mathrm{R}}^{(0)}+\mathfrak{L}_{\mathrm{R}}^{(1)}+\sum_{i=1}^3\mathfrak{L}_{i\mathrm{R}}^{(2)}+\ldots\label{elagragianhenrinf}
\end{equation} with
\begin{eqnarray}
\mathfrak{L}_{1\mathrm{R}}^{(2)}&=&-\frac{5\alpha\rm m^4 b}{96\pi^3}\int_{0}^{\infty}\frac{\rm ds}{\rm s^2}\rm  e^{-\rm s/b}\left(s\coth(\rm s)-1-\frac{\rm s^2}{3}\right)\nonumber\\&-&\frac{\alpha \rm m^4b^2}{16\pi^3}\int_{0}^\infty\rm \frac{ds}{s^3}\rm e^{-\rm s/b}\ln\left(\gamma\frac{s}{b}\right)\nonumber\\ &\times&\left[\rm s\coth(\rm s)+\frac{\rm s^2}{\sinh^2(\rm s)}-2\right]-\frac{\alpha\rm  m^4 b^2}{32\pi^3}\int_{0}^\infty\rm ds e^{-\rm s/b}\nonumber\\&\times&\int_{0}^{1} d\eta\left\{\tilde{\mathcal{Q}}_1(\rm s,\eta)-\tilde{\mathtt{Q}}_{1}(\rm s,\eta)-\tilde{\mathfrak{H}}_1(\rm s,\eta)\right\},\\
\mathfrak{L}_{2\mathrm{R}}^{(2)}&=&-\frac{5\alpha\rm m^4  b}{192\pi^3}\int_{0}^{\infty}\frac{\rm ds}{\rm s^2}\rm  e^{-\rm s/b}\left(s\coth(\rm s)-1-\frac{\rm s^2}{3}\right)\nonumber\\&-&\frac{\alpha\rm m^4 b^2}{32\pi^3}\int_{0}^\infty\rm \frac{d s}{s^3}\rm e^{-\rm s/b}\ln\left(\gamma\frac{s}{b}\right)\nonumber\\&\times&\left[3\rm s\coth(\rm s)+\frac{\rm s^2}{\sinh^2(\rm s)}-4-\frac{2\rm s^2}{3}\right]-\frac{\alpha\rm m^4 b^2}{32\pi^3}\int_{0}^\infty\rm ds\nonumber\\&\times&\rm e^{-\rm s/b}\int_{0}^{1} d\eta\left\{\tilde{\mathcal{Q}}_2(\rm s,\eta)-\tilde{\mathtt{Q}}_{2}(\rm s,\eta)-\tilde{\mathfrak{H}}_2(\rm s,\eta)\right\},\\
\mathfrak{L}_{3\mathrm{R}}^{(2)}&=&-\frac{15\alpha\rm m^4  b}{192\pi^3}\int_{0}^{\infty}\frac{\rm ds}{\rm s^2}\rm  e^{-\rm s/b}\left(s\coth(\rm s)-1-\frac{\rm s^2}{3}\right)\nonumber\\&-&\frac{\alpha\rm m^4 b^2}{32\pi^3}\int_{0}^\infty\rm \frac{d s}{s^3}\rm e^{-\rm s/b}\ln\left(\gamma\frac{s}{b}\right)\nonumber\\&\times&\left[\rm s\coth(\rm s)+\frac{3\rm s^2}{\sinh^2(\rm s)}-4+\frac{2\rm s^2}{3}\right]-\frac{\alpha\rm m^4 b^2}{32\pi^3}\int_{0}^\infty\rm ds\nonumber\\&\times&\rm e^{-\rm s/b}\int_{0}^{1} d\eta\left\{\tilde{\mathcal{Q}}_3(\rm s,\eta)-\tilde{\mathtt{Q}}_{3}(\rm s,\eta)-\tilde{\mathfrak{H}}_3(\rm s,\eta)\right\}\label{eeqww}
\end{eqnarray}
where we have replaced the integration variable $\tau$ by
$\rm s,$ and $\rm b=B/B_c.$ Here we have used the function $\mathfrak{f}_i(\rm s,\eta)\equiv\left\{\tilde{\mathcal{Q}}_i(\rm s,\eta)-\tilde{\mathtt{Q}}_{i}(\rm s,\eta)-\tilde{\mathfrak{H}}_i(\rm s,\eta)\right\}$ with  $\tilde{\mathcal{Q}}_{i}(\rm s,\eta)=\mathcal{Q}_{i}(\rm s,\eta)/(\rm eB)^3,$  $\tilde{\mathtt{Q}}_{i}(\rm s,\eta)=\mathtt{Q}_{i}(\rm s,\eta)/(\rm eB)^3$ and $\tilde{\mathfrak{H}}_{i}(\rm s,\eta)=\mathfrak{H}_{i}(\rm s,\eta)/(\rm eB)^3.$

\subsection{Asymptotic behavior at large magnetic field strength}

In the asymptotic region of superstrong magnetic field $\rm b\gg1,$  Eq. (\ref{1looprenor})  behaves like
\cite{ritus,dittrich1}
\begin{equation}
\mathfrak{L}_{\mathrm{R}}^{(1)}(\rm b)\simeq\frac{\rm m^4 b^2 }{24 \pi^2 \rm
}\left\{\ln\left(\frac{\rm b}{\gamma \pi}\right)+\frac{6}{\pi^2}\zeta^{\prime}(2)\right\},\label{1looprenorapp}
\end{equation} where $6\pi^{-2}\zeta^{\prime}(2)=-0.5699610\ldots$ with $\zeta(x)$  the Riemann zeta-function. 

For the same magnetic field regime the leading term of  $\mathfrak{L}_{i\mathrm{R}}^{(2)}$  is  derived  in  Appendix \ref{appendB}. 
\begin{eqnarray}
\begin{array}{c}
\displaystyle
\mathfrak{L}_{1\mathrm{R}}^{(2)}\approx-\frac{\alpha\rm m^4 b^2 }{16\pi^3}\mathcal{N}_1,\\ \\ \displaystyle
\mathfrak{L}_{2\mathrm{R}}^{(2)}\approx\frac{\alpha\rm m^4 b^2}{32\pi^3}\left[\mathcal{N}_2\ln\left(\frac{\rm b}{\gamma\pi}\right)-\frac{1}{3}\ln^2\left(\frac{\rm b}{\gamma\pi}\right)+\mathcal{A}\right],\\ \\ \displaystyle
\mathfrak{L}_{3\mathrm{R}}^{(3)}\approx\frac{\alpha\rm m^4 b^2}{32\pi^3}\left[\mathcal{N}_3\ln\left(\frac{\rm b}{\gamma\pi}\right)+\frac{1}{3}\ln^2\left(\frac{\rm b}{\gamma\pi}\right)+\mathcal{B}\right].
\end{array}\label{leading1}
\end{eqnarray} These expressions  are calculated with accuracy of terms  decreasing with $\rm b$ like $\sim \rm b^{-1}\ln(b)$ and faster. Here  the numerical constants are $\mathcal{N}_{1}=1.25,$  $\mathcal{N}_{2}=\frac{1}{3}-\frac{4\zeta^{\prime}(2)}{\pi^2},$  $\mathcal{N}_{3}=\frac{2}{3}+\frac{4\zeta^{\prime}(2)}{\pi^2},$ $\mathcal{A}=4.21$ and $\mathcal{B}=0.69.$ Note $\mathcal{N}_2+\mathcal{N}_3=1.$   

Taking all this into account, the asymptotic behavior of the full two-loop term is
\begin{eqnarray}
\mathfrak{L}_{\mathrm{R}}^{(2)}=\sum_{i=1}^3\mathfrak{L}_{i\mathrm{R}}^{(2)}\approx\frac{\alpha\rm m^4 b^2}{32\pi^3}\left[\ln\left(\frac{\rm b}{\gamma\pi}\right)+2.4\right],
\end{eqnarray} which coincides with the  results reported in references
\cite{ritus,ritus1,dittrich1}

\section{Vacuum magnetic properties  in a  superstrong magnetic field \label{sec:magnetvacuum}}

\subsection{Role of the photon polarization modes on  the vacuum magnetization}

\begin{figure}[!htbp]
\begin{center}
\includegraphics[width=3in]{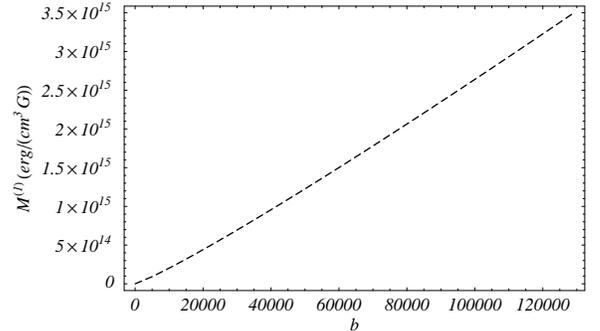}
\caption{\label{fig:mb1} One-loop contribution  to the vacuum magnetization density  with regard to the
external field.}
\end{center}
\end{figure}

In  presence of an  external magnetic field, the zero-point vacuum
energy $ \mathscr{E}_{\mathrm{vac}}$ is  modified by the  interaction  between $\textbf{B}$ and
the virtual QED-particles. The  latter is determined by the
effective potential coming from the  quantum-corrections to the
Maxwell Lagrangian which is also contained within the finite
temperature formalism. According to Eq. (\ref{elagragianhenrinf}) it is  expressed as
\begin{equation}
\mathscr{E}_{\mathrm{vac}}=-\mathfrak{L}_{\mathrm{R}}^{(1)}-\sum_{i=1}^3\mathfrak{L}_{i\mathrm{R}}^{(2)}+\ldots 
\end{equation} Consequently the vacuum  acquires a non trivial magnetization
$\mathrm{M}_{\mathrm{vac}}=-\partial\mathscr{E}_{\mathrm{vac}}/\partial\rm
\textbf{B}$ induced by the external magnetic field. In what follows
we will write
\begin{equation}
\mathrm{M}_{\mathrm{vac}}=\mathrm{M}_{\mathrm{vac}}^{(1)}+\mathrm{M}_{\mathrm{vac}}^{(2)}+\ldots 
\end{equation} in correspondence with the loop-term
$\mathfrak{L}^{(i)}_{\mathrm{R}}.$ In this sense, the one-loop
contribution at very large magnetic field $\rm b\gg1$ can be
computed  by means of Eq. (\ref{1looprenorapp}) and   gives:
\begin{equation}
\begin{array}{c}
\rm M^{(1)}=\frac{\partial\mathfrak{L}_{\mathrm{R}}^{(1)}}{\partial
\rm B}\approx\frac{\rm m^4 b}{24\pi^2 \rm B_c}\left[2\ln\left(\frac{\rm b}{\gamma\pi}\right)+1+\frac{12\zeta^{\prime}(2)}{\pi^2}\right] 
\end{array}.\label{magnetization1loop}
\end{equation}
The  dependence on the external field is  shown in FIG.
\ref{fig:mb1}. The data depicted  is obtained in the field
interval $10 \leq \rm b \leq 10^2\cdot 3\pi/\alpha$. Within this
approximation,  we find that the vacuum  reacts  paramagnetically and
has a nonlinear dependence on the external field (see Eq. (\ref{magnetization1loop})).

The two-loop correction is given by $\rm
M^{(2)}=\sum_{i=1}^3M_{i}^{(2)}$  where  $\rm
M_{i}^{(2)}=\partial\mathfrak{L}_{i\mathrm{R}}^{(2)}/\partial\textrm{B}$
is  the contribution corresponding to a  photon propagation
mode. Making use of Eqs. (\ref{leading1}) we find
\begin{eqnarray}
\mathrm{M}_1^{(2)}&\approx&-\frac{\alpha\rm m^4b}{8 \pi^3 \rm B_c}\mathcal{N}_1,\\ 
\mathrm{M}_2^{(2)}&\approx&-\frac{\alpha\rm m^4b}{32 \pi^3 \rm B_c}\left[\frac{2}{3}\ln^2\left(\frac{\rm b}{\gamma \pi}\right)+\frac{8\zeta^{\prime}(2)}{\pi^2}\ln\left(\frac{\rm b}{\gamma \pi}\right)\right.\nonumber\\&&-\left.\mathcal{N}_2-2\mathcal{A}\frac{}{}\right],\\  \mathrm{M}_3^{(2)}&\approx&\frac{\alpha\rm m^4b}{32 \pi^3 \rm B_c}\left[\frac{2}{3}\ln^2\left(\frac{\rm b}{\gamma \pi}\right)+\left(2+\frac{8\zeta^{\prime}(2)}{\pi^2}\right)\ln\left(\frac{\rm b}{\gamma \pi}\right)\right.\nonumber\\&&+\left.\mathcal{N}_3+2\mathcal{B}\frac{}{}\right].
\label{leadingmag1}
\end{eqnarray}

\begin{figure}[tp]
\begin{center}
\includegraphics[width=3in]{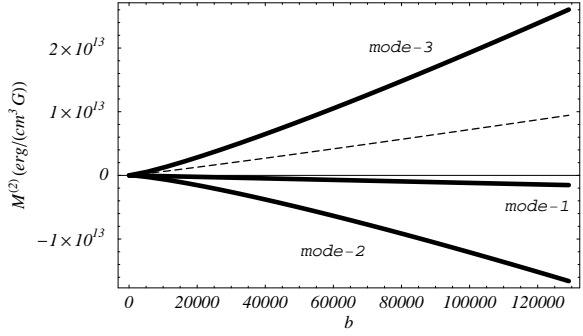}
\caption{\label{fig:mb3} Contribution of the vacuum polarization
eigenvalues to the  vacuum magnetization density  with regard to the
external field.  Here $10 <\rm b< 10^2\cdot 3\pi/\alpha.$  The
dashed line represents the complete two-loop contribution  to the
vacuum.}
\end{center}
\end{figure}

According to these results,  in a superstrong magnetic field 
limit, $\mathrm{M}_1^{(2)}<0$ and $\mathrm{M}_2^{(2)}<0$
behave diamagnetically whereas  $\mathrm{M}_3^{(2)}>0$ is purely
paramagnetic (see Fig. \ref{fig:mb3}). Within  the range of magnetic
field values for which  the photon  anomalous magnetic moment is
defined $(10^{14}\rm G\lesssim\vert
\textbf{B}\vert\lesssim10^{15}\rm G)$ the  vacuum magnetization
density is $\rm M_2^{(2)}\sim-10^{9}erg/(cm^3G).$  Moreover, while
$\mathrm{M}_1^{(2)}$ depends linearly on $\rm b,$ the  contributions
of the second and third propagation mode depend logarithmically on
the external field. Note that the leading behavior of the complete
two-loop contribution is
\begin{equation}
\mathrm{M}^{(2)}\approx\frac{\alpha\rm m^4 b}{32\pi^3\rm
B_c}\left[2\ln\left(\frac{\rm b}{\gamma\pi}\right)+5.8\right]>0
\end{equation} which points out a dominance of the third mode.

As it was expected  $\rm M^{(1)}/M^{(2)}\sim\alpha^{-1}.$ This ratio
is  also manifested between the  corresponding  magnetic susceptibilities ($\mathtt{X}^{(i)}=\partial
\mathrm{M}^{(i)}/\partial \mathrm{B}$). Note that
\begin{eqnarray}
\begin{array}{c}\displaystyle
\mathtt{X}^{(1)}\approx\frac{\rm m^4}{24 \pi^2 \rm
\rm B_c^2}\left[2\ln\left(\frac{\rm b}{\gamma\pi}\right)+1.86\right]>0,\\ \\ \displaystyle
\mathtt{X}^{(2)}\approx\frac{\alpha \rm m^4 }{32 \pi^3 \rm \rm
B_c^2}\left[2\ln\left(\frac{\rm b}{\gamma\pi}\right)+7.8\right]>0.
\end{array}
\end{eqnarray} For magnetic fields $\rm b\sim 10^5$ corresponding to  $\rm \vert\textbf{B}\vert\sim 10^{18}G,$  the
magnetic susceptibility reaches   values the order of
$\mathtt{X}^{(1)}\sim10^{-4}\rm  erg/(cm^3 G^2)$ which  exceeds the 
values of many laboratory materials, for example  Aluminum ($\mathtt{X}_{\mathrm{Al}}=2.2\cdot 10^{-5}\rm erg/(cm^3 G^2)$).

\subsection{Transverse pressures}

Because of  the anisotropy generated by $\bf B$ a magnetized vacuum
exerts two different pressure components  \cite{hugo4, Elizabeth1}.  One of
them is positive
$(\mathrm{P}_\parallel=-\mathscr{E}_{\mathrm{vac}})$ and along
$\textbf{B},$ whereas the remaining is  transverse to the external
field direction
($\mathrm{P}_\perp=-\mathscr{E}_{\mathrm{vac}}-\mathrm{M}\vert\textbf{B}\vert$).
For $\rm b\sim 1$  the  latter  acquires  negative values.

At very large magnetic  fields ($\rm b\gg1$) the one-loop
approximation of  $\mathrm{P}_\perp$ can be computed by making use of
Eq. (\ref{1looprenorapp}) and Eq. (\ref{magnetization1loop}).  In
fact
\begin{equation}
\rm P_\perp^{(1)}\approx-\frac{\rm m^4\rm b^2}{24 \pi^2}\left[\ln\left(\frac{\rm b}{\gamma \pi}\right)+1+\frac{6\zeta^\prime(2)}{\pi^2}\right]<0.
\end{equation} Therefore,  at  asymptotically  large  values of the external field,
the interaction between $\bf B$ and the virtual  electron-positron pairs  generates a negative pressure which would tend to shrink inserted matter in the plane transverse  to  $\textbf{B}.$

Again, the two-loop contribution  can be written as the sum of the
corresponding  terms due to the  vacuum polarization modes
\begin{equation}
 \mathrm{P}_{\perp}^{(2)}=\sum_{i=1}^3\mathrm{P}_{\perp i}^{(2)}. 
\end{equation}
According to Eqs. (\ref{leading1}) and Eqs.
(\ref{leadingmag1}) they read:
\begin{eqnarray}
\mathrm{P}_{\perp1}^{(2)}&\approx&\frac{\alpha \rm m^4\rm b^2}{16 \pi^3}\mathcal{N}_1>0,\\  \mathrm{P}_{\perp2}^{(2)}&\approx&\frac{\alpha \rm m^4\rm b^2}{32 \pi^3}\left[\frac{1}{3}\ln^2\left(\frac{\rm b}{\gamma\pi}\right)+\left(\mathcal{N}_2+\frac{8\zeta^{\prime}(2)}{\pi^2}\right)\ln\left(\frac{\rm b}{\gamma\pi}\right)\right.\nonumber\\ &&-\left.\mathcal{N}_2-\mathcal{A}\frac{}{}\right]>0,\\ 
\mathrm{P}_{\perp3}^{(2)}&\approx&-\frac{\alpha \rm m^4\rm b^2}{32 \pi^3}\left[\frac{1}{3}\ln^2\left(\frac{\rm b}{\gamma\pi}\right)+\left(2-\mathcal{N}_3+\frac{8\zeta^{\prime}(2)}{\pi^2}\right)\right.\nonumber\\ &&\left.\times\ln\left(\frac{\rm b}{\gamma\pi}\right)+\mathcal{N}_3+\mathcal{B}\right]<0,
\end{eqnarray} 

\begin{figure}[tp]
\includegraphics[width=3in]{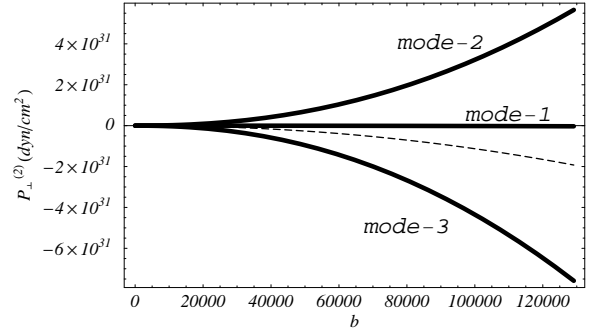}
\caption{\label{fig:mb87} Contribution of the vacuum polarization
eigenvalues to the vacuum transverse pressure  density with regard
to the external field.  Here $10 <\rm b< 10^2\cdot 3\pi/\alpha.$ The
dashed line represents the complete two-loop contribution.}
\end{figure}
with the complete two-loop term given by 
\begin{equation}
\mathrm{P}_{\perp}^{(2)}\approx -\frac{\alpha\rm m^4
b^2}{32\pi^3}\left[\ln\left(\frac{\rm
b}{\gamma\pi}\right)+3.4\right]<0. 
\end{equation}For $\rm b\sim 10 ^5,$ corresponding to magnetic fields  $\rm
B\sim10^{18}G,$ the transverse pressure generated by  the first and
second polarization mode is  positive and reaches values of the order
$\sim10^{30}\rm dyn/cm^2$ and  $\sim 10^{31}\rm dyn/cm^2,$
respectively (see Fig. \ref{fig:mb87}). In contrast, the contribution given by the  third
mode is negative with   $\mathrm{P}_{\perp3}^{(2)}\sim-10^{31}\rm
dyn/cm^2.$ In the same context
$\mathrm{P}_{\perp}^{(2)}\sim-10^{31}\rm dyn/cm^2.$ The one-loop
contribution is correspondingly even of the order of  $\mathrm{P}_{\perp}^{(1)}\sim
\alpha^{-1}\mathrm{P}_{\perp}^{(2)}\sim-10^{33}\rm dyn/cm^2.$

\section{Conclusion}

In the first part of this work we showed that an infrared photon propagating in
a strongly magnetized vacuum $(10<\rm b\ll 3\pi/\alpha)$, seems to exhibit a
nonzero vector anomalous magnetic moment [see Eq. (\ref{goddeft})].  We have
pointed out that  this quantity is a signal of the Lorentz symmetry
breaking due to  the presence of an  external magnetic field.  In addition, we have shown that $\boldsymbol{m}_\gamma$  arises due to the
interaction between virtual electron-positron quantum pairs with the
external magnetic field and its existence is closely related to the
gauge invariance. In this context,  we showed that $\boldsymbol{m}_\gamma$ can be
decomposed into two orthogonal components in correspondence with the
cylindrical symmetry imposed by the external field. These components
are opposite in sign, and only the one  along $\bf B$ is conserved.
We remarked  that the photon paramagnetism (analyzed in \cite{hugoe})
is only associated with a kind of photon magnetization rather than   a magnetic moment. 

In the  last sections of this  work we discussed the  effect of the
vacuum polarization tensor, which modifies the zero-point  energy of
the vacuum even in the absence of electromagnetic waves. In the limit of  a
superstrong  magnetic field, the two-loop contribution
of the magnetization density corresponding to the  second and third
propagation mode depends  nonlinearly on the external magnetic
field and their behavior is diamagnetic and paramagnetic,
respectively. On the other hand, the contribution coming from the
first mode is diamagnetic and depends linear on  $\rm B$.  We have
seen that for very large magnetic field the contribution due to the
third mode strongly dominate the analyzed quantities. In this
magnetic field regime the latter exerts  a negative transverse pressures to the external field. On
the contrary those contributions coming from the first and second
virtual mode are  positive.

We want to point out that, although the decomposition of the  two-loop term in the
Euler-Heisenberg Lagrangian was  considered in a   magnetic 
background it remains valid also for the electric  case. This
fact should  allow to study  the role of the vacuum polarization
modes in  electron-positron production in a superstrong electric
field. A detailed analysis of this issue will be presented in a
forthcoming work.

\begin{acknowledgments}
The author expresses its deep gratitude to
professor A. Shabad for  valuable suggestions and advice. He also
thanks H. P\'erez Rojas for several discussions and important
remarks, to  Felipe J. Llanes-Estrada, F. Hebenstreit  and Danny Martinez-Pedrera for
a critical reading and comments to improve the first version of this
manuscript. The author wants  to extend his gratitude to professor
Reinhard Alkofer for helping to  support this research. This work
has been supported by the Doktoratskolleg ``Hadrons in Vacuum,
Nuclei and Stars''  of the Austrian Science Fund (FWF) under
contract W1203-N08.
\end{acknowledgments}

\appendix
\section{\label{hfgt}}

The  term   $\rm z_2$ can be written as
\begin{eqnarray}
\rm z_2&=&\frac{1}{2\Delta}[\textbf{B}\times\textbf{k}]^2=\textbf{k}^2-\frac{(\textbf{B}\cdot\textbf{k})^2}{2\Delta}
\end{eqnarray} and its  substitution  into  Eq. (\ref{comics}) leads to:
\begin{equation}
\omega^2=\frac{1}{1+\varrho(\textrm{b})}\left(\textbf{k}^2-\varrho(\textrm{b})\frac{(\textbf{B}\cdot\textbf{k})^2}{2\Delta}\right).
\end{equation}
Defining the photon magnetization by means of the
relation $\boldsymbol{\mu}_\gamma=-\partial\omega/\partial \textbf{B}$ we find 
\begin{eqnarray}
\boldsymbol{\mu}_\gamma=-\frac{\partial\omega}{\partial
\textbf{B}}&=&\frac{1}{2\omega(1+\varrho(\textrm{b}))^2}\left[\textbf{k}^2+(1+2\varrho(\textrm{b}))\frac{(\textbf{B}\cdot\textbf{k})^2}{2\Delta}\right]\nonumber\\
&\times&\frac{\textbf{B}}{2\Delta}-\frac{\varrho(\textrm{b})}{2\omega(1+\varrho(\textrm{b}))}\frac{(
\textbf{B}\cdot\textbf{k})}{\Delta}\textbf{k}.
\end{eqnarray} For $\textbf{B}_\perp\to0$ we obtain
\begin{eqnarray}\label{gmgparallel}
\begin{array}{c}\displaystyle
\boldsymbol{\mu}^\perp_\gamma=\left.\frac{\partial \omega}{\partial
\textbf{B}_\perp}\right\vert_{\textbf{B}_\perp=0}=-\frac{\varrho(\textrm{b})}{\omega(1+\varrho(\textrm{b}))\vert\rm
B_z\vert}\rm k_\parallel\textbf{k}_\perp,\\ \\ \displaystyle
\boldsymbol{\mu}^\parallel_\gamma=\left.\frac{\partial \omega}{\partial
\textbf{B}_\parallel}\right\vert_{\textbf{B}_\perp=0}=\frac{\varrho(\textrm{b})\rm
k_\perp^2}{2\omega(1+\varrho(\textrm{b}))^2\vert\rm
B_z\vert}\textbf{n}_\parallel.
\end{array}
\end{eqnarray}

Note that for  magnetic field strength  $10<\rm b \ll 3\pi/\alpha$ one should
treat $\varrho\ll1.$  As consequence Eq. (\ref{gmgparallel}) becomes
\begin{eqnarray}
\boldsymbol{\mu}^\perp_\gamma=-\mathfrak{g}\frac{\rm
e}{\textrm{m}_0}\textbf{n}_\perp \rm \cos\phi\ \ \mathrm{and}\ \
\boldsymbol{\mu}^\parallel_\gamma=\mathfrak{g}\frac{\rm
e}{2\textrm{m}_0}\textbf{n}_\parallel \rm \sin\phi.\label{mmaprox}
\end{eqnarray}

\subsection{Relation between $\boldsymbol{\mu}_\gamma$ and $\boldsymbol{m}_\gamma$ }
According to these expressions   we can write
\begin{equation}
\boldsymbol{\mu}_\gamma=\boldsymbol{m}_\gamma+\frac{\partial\boldsymbol{m}_\gamma}{\partial \textbf{B}}\cdot\textbf{B}.\label{deribvfg}
\end{equation} The second term in  Eq. (\ref{deribvfg}) is a vector orthogonal 
to the external field. It can be expressed as
\begin{equation}
\frac{\partial\boldsymbol{m}_\gamma}{\partial \textbf{B}}\cdot\textbf{B}=\mathfrak{g}\frac{\rm e}{2\rm m}\mathbf{n}_\perp\cos\phi=\mathfrak{g}\frac{\rm e}{2\rm m}(\mathbf{n}\cdot\mathbf{n}_\parallel)\left(\mathbf{n}_\parallel\times \mathbf{s}_\gamma\right)\label{thgfd}
\end{equation}By substituting  Eq. (\ref{goddeft}) and Eq. (\ref{thgfd}) in Eq. (\ref{deribvfg}) we obtain
\begin{equation}
\boldsymbol{\mu}_\gamma=\mathfrak{g}\frac{\rm e}{2\rm m}\left[\mathbf{n}+(\mathbf{n}\cdot\mathbf{n}_\parallel)\mathbf{n}_\parallel\right]\times \mathbf{s}_\gamma.
\end{equation}

Substituting  Eq. (\ref{deribvfg}) in Eq. (\ref{spinsas}) gives 
\begin{equation}
\mathscr{U}=-\boldsymbol{m}_\gamma\cdot\textbf{B}=-\boldsymbol{\mu}_\gamma\cdot\textbf{B}+\mathfrak{g}\frac{\rm e}{2\rm m}\cos\phi \textbf{n}_\perp\cdot\textbf{B}\label{sdkafopas}
\end{equation}Because of the fact that  the second term on  the right-hand side is projected out by the scalar
product  $( \textbf{n}_\perp\cdot\textbf{B}=0),$ the physical consequences
of $\boldsymbol{\mu}_\gamma$  might be analyzed by means of  $\boldsymbol{m}_\gamma.$ So,
in a good approximation
\begin{equation}
\boldsymbol{\mu}_\gamma\simeq \boldsymbol{m}_\gamma.
\end{equation} the remaining transversal component involved in $\boldsymbol{m}_\gamma$  is not relevant because  is projected out by the scalar product $\boldsymbol{m}_\gamma\cdot\textbf{B}$.

\subsection{$\boldsymbol{\mu}^\parallel_\gamma$ for $\rm b\geq 3\pi/\alpha$ }
Note, in addition, that  for  purely  perpendicular propagation,   Eq. (\ref{gmgparallel})  reduces to: 
\begin{equation}
\mu^\parallel_\gamma=\mathfrak{g}\frac{\rm e}{2\rm m}f[\rm k_\perp](1+\varrho(\rm b))^{-5/2}. 
\end{equation} In particular for $\rm b\to\infty$ 
\[\mu^\parallel_\gamma\approx\mathfrak{g}\frac{\rm e}{2\rm m}f[\rm k_\perp](\varrho(\rm b))^{-5/2}.\]

\section{\label{appendB}}

The aim of this appendix  is to  find  the asymptotic behavior
of $\mathfrak{L}_{i\mathrm{R}}^{(2)}$ for  $\rm b\gg1.$ For this purpose, it is     convenient to write
\begin{equation}
\mathfrak{L}_{i\mathrm{R}}^{(2)}=\mathfrak{L}_{i\mathrm{R}}^{(2)1}+\mathfrak{L}_{i\mathrm{R}}^{(2)2}\label{sppliting}
\end{equation} with  the first term  being given by 
\begin{widetext}
\begin{eqnarray}\label{2looppart11}
\begin{array}{c}\displaystyle
\mathfrak{L}_{1\mathrm{R}}^{(2)1}=-\frac{5\alpha\rm m^2 }{12\pi}\frac{\partial\mathfrak{L}_\mathrm{R}^{(1)}}{\partial \rm m^2}-\frac{\alpha\rm m^2 b^2}{16\pi^3}\Sigma_1-\frac{\alpha \rm m^2 }{2\pi}\ln\left(\frac{\rm b}{\gamma\pi}\right)\frac{\partial\mathfrak{L}_\mathrm{R}^{(1)}}{\partial \rm m^2},\\\displaystyle
\mathfrak{L}_{2\mathrm{R}}^{(2)1}=-\frac{5\alpha\rm m^2 }{24\pi}\frac{\partial\mathfrak{L}_\mathrm{R}^{(1)}}{\partial \rm m^2}-\frac{\alpha\rm  m^4 b^2}{32\pi^3}\Sigma_2-\frac{\alpha}{2\pi}\ln\left(\frac{\rm b}{\gamma\pi}\right)\left\{\mathfrak{L}_\mathrm{R}^{(1)}+\frac{\rm m^2}{2}\frac{\partial\mathfrak{L}_\mathrm{R}^{(1)}}{\partial \rm m^2}\right\},\\\displaystyle
\mathfrak{L}_{3\mathrm{R}}^{(2)1}=-\frac{15\alpha\rm m^2 }{24\pi}\frac{\partial\mathfrak{L}_\mathrm{R}^{(1)}}{\partial \rm m^2}-\frac{\alpha\rm  m^4 b^2}{32\pi^3}\Sigma_3+\frac{\alpha}{2\pi}\ln\left(\frac{\rm b}{\gamma\pi}\right)\left\{\mathfrak{L}_\mathrm{R}^{(1)}-\frac{3\rm m^2}{2}\frac{\partial\mathfrak{L}_\mathrm{R}^{(1)}}{\partial \rm m^2}\right\}, 
\end{array}
\end{eqnarray}
with
\begin{eqnarray}
\begin{array}{c}\displaystyle
\Sigma_1=\int_{0}^\infty\frac{\rm ds}{\rm s^3}\rm e^{-s/b} \ln\left(\frac{\rm s}{\pi}\right)\left(\rm s\coth(s)+\frac{s^2}{\sinh^2(s)}-2\right),\ \ \displaystyle
\Sigma_2=\int_{0}^\infty\frac{\rm ds}{\rm s^3}\rm e^{-s/b} \ln\left(\frac{\rm s}{\pi}\right)\left(\rm 3s\coth(s)+\frac{s^2}{\sinh^2(s)}-4-\frac{2\rm s^2}{3}\right),\\\displaystyle
\Sigma_3=\int_{0}^\infty\frac{\rm ds}{\rm s^3}\rm e^{-s/b} \ln\left(\frac{\rm s}{\pi}\right)\left(\rm s\coth(s)+3\frac{s^2}{\sinh^2(s)}-4+\frac{2\rm s^2}{3}\right). 
\end{array}
\end{eqnarray} 
The second term in   Eq. (\ref{sppliting}) is defined as:
\begin{equation}
\mathfrak{L}_i^{(2)2}=-\frac{\alpha\rm m^4 b^2}{32\pi^3}\mathrm{G}_i(\rm b)\ \ \mathrm{with}\ \ \mathrm{G}_i(\rm b)=\int_{0}^1\rm d\eta\int_{0}^\infty\rm ds e^{-\rm s/b}\mathfrak{f}_i(\rm s,\eta)\label{secondpartoftwoloopterm}
\end{equation}with  $
\mathfrak{f}_i(\rm s,\eta)\equiv\left\{\tilde{\mathcal{Q}}_i(\rm s,\eta)-\tilde{\mathtt{Q}}_{i}(\rm s,\eta)-\tilde{\mathfrak{H}}_i(\rm s,\eta)\right\}.$

\subsection{Leading  behavior of $\mathfrak{L}_{i\mathrm{R}}^{(2)1}$ in an  asymptotically large magnetic field }

In order to  determine  the leading asymptotic-magnetic field term of
$\mathfrak{L}_{i\mathrm{R}}^{(2)1},$  we  substitute  Eq. (\ref{1looprenorapp}) into Eqs.
(\ref{2looppart11}), which  gives:
\begin{eqnarray}\label{firtpart2loop1}
\begin{array}{c}\displaystyle
\mathfrak{L}_{1\mathrm{R}}^{(2)1}\simeq\frac{\alpha\rm m^4 b^2 }{16\pi^3}\left[\frac{1}{3}\ln\left(\frac{\rm b}{\gamma\pi}\right)+\frac{5}{18}-\Sigma_1\right],\ \  \displaystyle
\mathfrak{L}_{2\mathrm{R}}^{(2)1}\simeq\frac{\alpha\rm m^4 b^2}{32\pi^3}\left[\left(\frac{1}{3}-\frac{4\zeta^{\prime}(2)}{\pi^2}\right)\ln\left(\frac{\rm b}{\gamma\pi}\right)-\frac{2}{3}\ln^2\left(\frac{\rm b}{\gamma\pi}\right)+\frac{5}{18}-\Sigma_2\right],\\ \\ \displaystyle
\mathfrak{L}_{2\mathrm{R}}^{(3)1}\simeq\frac{\alpha\rm m^4 b^2}{32\pi^3}\left[\left(1+\frac{4\zeta^{\prime}(2)}{\pi^2}\right)\ln\left(\frac{\rm b}{\gamma\pi}\right)+\frac{2}{3}\ln^2\left(\frac{\rm b}{\gamma\pi}\right)+\frac{5}{6}+\Sigma_3\right].
\end{array}
\end{eqnarray}
\end{widetext}

Note that $\Sigma_2$ and $\Sigma_3$ can be expressed as
\begin{eqnarray}
\begin{array}{c}\displaystyle
\Sigma_2=\Sigma_1+2\Sigma,\ \ \displaystyle
\Sigma_3=\Sigma_1-2\Sigma-2\rm b\frac{\rm d\Sigma}{\rm d\rm b},
\end{array}\label{sigmaintegr}
\end{eqnarray} with
\begin{eqnarray}
\Sigma=\int_{0}^{\infty}\frac{\rm ds}{\rm s^3}\ln\left(\frac{\rm s}{\pi}\right)\rm e^{-\rm s/b}\left[\rm s \coth(s)-1-\frac{s^2}{3}\right].\label{sigmaintegr1}
\end{eqnarray} To derive the second expression in Eq. (\ref{sigmaintegr}) we
have  used the identity
\begin{eqnarray}\label{identiy1}
\rm s\coth(s)&+&\frac{s^{2}}{\sinh^2(\rm s)}-2=\\&-&\rm s^3\frac{\rm d}{\rm d s}\left[\frac{1}{\rm s^2}\left(\rm s\coth(s)-1-\frac{\rm s^2}{3}\right)\right]\nonumber
\end{eqnarray} and an integration by parts.

Note  that  $\Sigma_1$ converges even without the
exponential  factor which  approaches to $1$ for $\rm b\to \infty$.  By using MATHEMATICA code  we
find $\Sigma_1\simeq0.19.$

$\Sigma$ does not involve singularities in the integrands at $\rm
s=0,$ but  would diverge at $\rm s\to \infty$ if one sets the
limiting value $\exp(-\rm s/b)=1.$ For that reason we  divide
the integration domain into two parts:
\begin{equation}
\Sigma=\Sigma^{(\mathrm{L})}+\Sigma^{(\mathrm{H})}
\end{equation} with  $\Sigma_1\simeq0.19$
\begin{eqnarray}
\begin{array}{c}\displaystyle
\Sigma^{(\mathrm{L})}=\int_{0}^{\rm T}\frac{\rm ds}{\rm s^3}\ln\left(\frac{\rm s}{\pi}\right)\rm e^{-\rm s/b}\left[\rm s \coth(s)-1-\frac{s^2}{3}\right],\\ \\ \displaystyle
\Sigma^{(\mathrm{H})}=\int_{T}^{\infty}\frac{\rm ds}{\rm s^3}\ln\left(\frac{\rm s}{\pi}\right)\rm e^{-\rm s/b}\left[\rm s \coth(s)-1-\frac{s^2}{3}\right]
\end{array}
\end{eqnarray} and  $\rm T$ an arbitrary positive number.

Now, we can omit the exponential in $\Sigma_{\mathrm{L}}$ since  the
resulting integral converges anyway:
\begin{eqnarray}
\Sigma^{(\mathrm{L})}&\simeq&\int_{0}^{\rm T}\frac{\rm ds}{\rm s^3}\ln\left(\frac{\rm s}{\pi}\right)\left[\rm s \coth(s)-1-\frac{s^2}{3}\right].
\end{eqnarray}

Substantial simplification is  achieved  by  splitting the integrand of  $\Sigma_{\mathrm{H}}$ into its parts and neglecting the exponential factor $\exp\left(-\rm s/b\right)$ whenever is possible
\begin{eqnarray}
\Sigma^{(\mathrm{H})}&\simeq&\int_{T}^{\infty}\frac{\rm ds}{\rm s^2}\ln\left(\frac{\rm s}{\pi}\right)\coth(\rm s)-\int_{T}^{\infty}\frac{\rm ds}{\rm s^3}\ln\left(\frac{\rm s}{\pi}\right)\nonumber\\&-&\frac{1}{3}\int_{T}^{\infty}\frac{\rm ds}{\rm s}\ln\left(\frac{\rm s}{\pi}\right)\exp(\rm -s/b).
\end{eqnarray}For $\rm s\to\infty$ the leading term of  $\coth(\rm s)\simeq1$. Having this in mind, we  compute the integrals: 
\begin{eqnarray}
\begin{array}{c}\displaystyle
\int_{T}^{\infty}\frac{\rm ds}{\rm s^2}\ln\left(\frac{\rm s}{\pi}\right)\left.\cosh(\rm s)\right\vert_{\rm s\to\infty}=\frac{1+\ln\left(\frac{\rm T}{\pi}\right)}{\rm T},\\ \\ \displaystyle
\int_{T}^{\infty}\frac{\rm ds}{\rm s^3}\ln\left(\frac{\rm s}{\pi}\right)=\frac{1-2\ln\left(\frac{\rm T}{\pi}\right)}{4\rm T^2},
\end{array}
\end{eqnarray}and 
\begin{eqnarray}
\int_{T}^{\infty}\frac{\rm ds}{\rm s}\ln\left(\frac{\rm s}{\pi}\right)\rm e^{-\rm s/b}&\simeq&\frac{\pi^2}{12}+\frac{1}{2}\ln^2\left(\frac{\rm b}{\pi\gamma}\right)-\frac{1}{2}\ln^2\left(\frac{\rm T}{\pi\gamma}\right)\nonumber\\&-&\ln\left(\frac{\rm T}{\gamma}\right)\ln\left(\gamma\right)-\frac{1}{2}\ln^2\left(\frac{\rm \pi}{\gamma}\right).
\end{eqnarray} In the latter we have neglected  terms decreasing
as $\sim \rm T/b.$  Numerical calculation using MATHEMATICA  code
gives  $\rm T\simeq\pi$ by minimizing $\Sigma$ $(\rm d \Sigma/dT=0).$  Therefore,
\begin{equation}
 \Sigma\simeq-\frac{1}{6}\ln^2\left(\frac{\rm b}{\pi\gamma}\right)+0.28
\end{equation} and according to  Eq. (\ref{sigmaintegr}) 
\begin{equation}
\begin{array}{c}\displaystyle
\Sigma_1=0.19,\ \
\Sigma_2\simeq-\frac{1}{3}\ln^2\left(\frac{\rm b}{\pi\gamma}\right)+0.75,
\\\displaystyle \Sigma_3\simeq\frac{1}{3}\ln^2\left(\frac{\rm b}{\pi\gamma}\right)-0.04.
\end{array}\label{firstpart2loop11}
 \end{equation}

The substitution of the latter into  Eqs. (\ref{firtpart2loop1}) yields
\begin{eqnarray}
\begin{array}{c}
\mathfrak{L}_{1\mathrm{R}}^{(2)1}\simeq\frac{\alpha\rm m^4 b^2 }{16\pi^3}\left[\frac{1}{3}\ln\left(\frac{\rm b}{\gamma\pi}\right)+0.09\right],\\ \\
\mathfrak{L}_{2\mathrm{R}}^{(2)1}\simeq\frac{\alpha\rm m^4 b^2}{32\pi^3}\left[\mathcal{N}_2\ln\left(\frac{\rm b}{\gamma\pi}\right)-\frac{1}{3}\ln^2\left(\frac{\rm b}{\gamma\pi}\right)-0.47\right],\\ \\
\mathfrak{L}_{2\mathrm{R}}^{(3)1}\simeq\frac{\alpha\rm m^4 b^2}{32\pi^3}\left[\left(1+\frac{4\zeta^{\prime}(2)}{\pi^2}\right)\ln\left(\frac{\rm b}{\gamma\pi}\right)+\frac{1}{3}\ln^2\left(\frac{\rm b}{\gamma\pi}\right)+0.87\right].
\end{array}\nonumber
\end{eqnarray}

\subsection{Leading  behavior of $\mathfrak{L}_{i\mathrm{R}}^{(2)2}$ in an  asymptotically large magnetic field}

The asymptotic behavior of $\mathfrak{L}_{i\mathrm{R}}^{(2)2}$  is
obtained from the integral $\rm G_i.$ We  start our analysis by dividing the integration domain into two regions:
\begin{eqnarray}
\mathrm{G}_i=\int_{0}^{\rm T}\rm ds\int_{0}^{1}d\eta\ldots+\int_{\rm T}^{\infty}\rm ds\int_{0}^{1}d\eta\ldots
\end{eqnarray} with $\rm T>0$. We  denote the corresponding integrals by 
$\mathrm{G}_i^{(\rm L)}$  and  $\mathrm{G}_i^{(\rm H)},$
respectively. In the latter we replace the upper integration limit
over $\eta-$variable by a parameter $\eta_0=1-\rm T/s>0.$ In order to find  the behavior of  $\mathrm{G}_i^{(\rm L)}$ we  write
\begin{eqnarray}
\mathrm{G}_i^{(\rm L)}&=&\frac{1}{\rm b}\int_{0}^{\rm T}\rm d\rm s\int_{0}^{1}d\eta\mathfrak{f}_i\left(\rm \rm s,\eta\right)\exp(-\rm s/b)\\
&\simeq&\int_{0}^{\rm T}\rm d s\int_{0}^{1}d\eta\mathfrak{f}_i\left(\rm s,\eta\right)
\end{eqnarray} where we have set $\exp(\rm s/b)\simeq1$ since $\mathfrak{f}_i$ converges
within the corresponding integration domain. For $\rm T\to 0$
the behavior of the remaining  integrand  is
\begin{eqnarray}
\begin{array}{c}\displaystyle
\mathfrak{f}_1\left(\rm
s,\eta\right)\simeq\frac{1}{180}(\eta^2-13)\rm s,\ \  \mathfrak{f}_2\left(\rm
s,\eta\right)\simeq\frac{1}{270}(\eta^2-3)\rm s ,\\ \\  \displaystyle\mathfrak{f}_3\left(\rm
s,\eta\right)\simeq-\frac{1}{540}(\eta^2+27)\rm s.
\end{array}
\end{eqnarray}
The  integrals over $\rm s$ and $\eta$ are trivial to perform and give:
\begin{eqnarray}
\begin{array}{c}\displaystyle
\mathrm{G}_1^{(\rm L)}\approx-\frac{19}{540}\rm T^2,\ \ \mathrm{G}_2^{(\rm L)}\approx-\frac{2}{405}\rm T^2,\\ \\\displaystyle \mathrm{G}_3^{(\rm L)}\approx-\frac{41}{1620}\rm T^2. \end{array}\label{1lowsapproyt}
\end{eqnarray}

Let us consider now, the contributions coming from  the second
integration domain  which concerns to  $\mathrm{G}_i^{(\rm H)}$. In
order to this we consider the asymptotic expression of
$\mathfrak{f}_i(\rm s, \eta)$ for $\rm s\to \infty.$ First of all,
the asymptotic expansion of Eqs. (\ref{poleig4}) in powers of $\exp(-\rm s)$ and $\exp(\rm s\eta)$
produces an expansion of Eq. (\ref{poleig2}) in a sum of
contributions coming from the thresholds, the singular behavior in
the  threshold points originating from the divergencies of the $\rm
s-$integration in Eq. (\ref{poleig2}) near $\rm s=\infty$ as it was
developed in \cite{shabad1}. The leading terms in the expansion of
Eqs. (\ref{poleig4})   at $\rm
s\rightarrow\infty$ are
\begin{eqnarray}
\begin{array}{c}\displaystyle
\left.\left(\frac{\sigma_1(\rm s,\eta)}{\sinh
\rm s}\right)\right\vert_{\rm s\rightarrow
\infty}\simeq\frac{1+\eta}{4}\rm e^{-(1+\eta)\rm s}+\frac{1-\eta}{4}\rm e^{-(1-\eta)\rm s}\\ \\  \displaystyle
\left.\left(\frac{\sigma_2(\rm s,\eta)}{\sinh
\rm s}\right)\right\vert_{\rm s\rightarrow \infty
}\simeq\frac{1-\eta^2}{4},\\ \\ \displaystyle
\left.\left(\frac{\sigma_3(\rm s,\eta)}{\sinh
\rm s}\right)\right\vert_{\rm s\rightarrow
\infty}\simeq2\exp\left(-2\rm s\right).
\end{array}
\end{eqnarray} These expressions are in  correspondence with the lowest
threshold ($n=0$, $n^\prime=1$ or viceversa for $i=1$,
$n=n^\prime=1$ for $i=3$, and $n=n^\prime=0$ for $i=2$). The fact that 
$\rm M(\infty,\eta)\simeq1/2$  and   $\rm
N(\infty,\eta)\gg M(\infty,\eta)$  allow us to write
\begin{eqnarray}
\begin{array}{c}\displaystyle
\mathtt{V}(\rm s,\eta)\simeq-\frac{2}{\mathrm{N}^2(\rm s,\eta)}+\frac{2\ln\left[2\rm N(s,\eta)\right]}{\rm N^2(s,\eta)}\simeq\frac{2\ln\left[2\rm N\right]}{\rm N^2}\\ \\ \displaystyle
\mathtt{W}(\rm s,\eta)\simeq\frac{4}{\rm N(s,\eta)}-\frac{2\ln\left[2\rm N(s,\eta)\right]}{\rm N^2(s,\eta)}\simeq\frac{4}{\rm N(s,\eta)}
\end{array}
\end{eqnarray} with  $\mathtt{V}(s,\eta)+\mathtt{W}(s,\eta)\simeq 4\rm N^{-1}(s,\eta).$ Considering  only the terms which decrease most slowly as a function of  $\rm s,$ we find
\begin{eqnarray}\label{c11}
\begin{array}{c}\displaystyle
\left.\tilde{\mathcal{Q}}_{1}(\rm s,\eta)\right\vert_{\rm s\to \infty}\simeq\frac{4\rm e^{-(1-\eta)\rm s}}{(1+\eta)\rm s}+\frac{4\rm e^{-(1+\eta)\rm s}}{(1-\eta)\rm s},\\ \\ \displaystyle\left.\tilde{\mathcal{Q}}_{2}(\rm s,\eta)\right\vert_{\rm s\to\infty}\simeq\left.\mathcal{Q}_{1}(\rm s,\eta)\right\vert_{\rm s\to \infty},\\ \\ \displaystyle
 \left.\tilde{\mathcal{Q}}_{3}(\rm s,\eta)\right\vert_{\rm s\to\infty}\simeq\frac{8\exp(-2\rm s)}{(1-\eta^2)\rm s}.
\end{array}
\end{eqnarray} Additionally,  the most significant  terms  arising
from $\tilde{\mathtt{Q}}_i+\tilde{\mathfrak{H}}_i$ for $\rm s\to \infty$ are
\begin{eqnarray}\label{c16}
\begin{array}{c}\displaystyle
\left.\left[\tilde{\mathtt{Q}}_1(\rm s,\eta)+\tilde{\mathfrak{H}}_{1}(\rm s,\eta)\right]\right\vert_{\rm s\to\infty}\simeq-\frac{2}{3\rm s}+\frac{16\rm e^{-2\rm s}}{(1-\eta^2)\rm s},\\ \\ \displaystyle
\left.\left[\tilde{\mathtt{Q}}_2(\rm s,\eta)+\tilde{\mathfrak{H}}_{2}(\rm s,\eta)\right]\right\vert_{\rm s\to\infty}\simeq\frac{8\rm e^{-2\rm s}}{(1-\eta^2)\rm s},\\ \\ \displaystyle
\left.\left[\tilde{\mathtt{Q}}_3(\rm s,\eta)+\tilde{\mathfrak{H}}_{3}(\rm s,\eta)\right]\right\vert_{\rm s\to\infty}\simeq-\frac{1}{3\rm s}+\frac{24\rm e^{-2\rm s}}{(1-\eta^2)\rm s}\end{array}.
\end{eqnarray}The behavior of $\mathfrak{f}_i=\exp(-\rm b/s)(\tilde{\mathcal{Q}}_i-\tilde{\mathtt{Q}}_i-\tilde{\mathfrak{H}}_i)$ for $\rm s\to \infty$ is find out by considering Eqs. (\ref{c11}-\ref{c16}). In fact
\begin{eqnarray}
\begin{array}{c}\displaystyle
\left.\mathfrak{f}_{1}\right\vert_{\rm s\to\infty}\simeq\frac{4\rm e^{-(1-\eta)\rm s}}{(1+\eta)\rm s}+\frac{4\rm e^{-(1+\eta)\rm s}}{(1-\eta)\rm s}-\frac{16\rm e^{-2\rm s}}{(1-\eta^2)\rm s}+\frac{2}{3\rm s},\\ \\ \displaystyle
\left.\mathfrak{f}_{2}\right\vert_{\rm s\to\infty}\simeq\frac{4\rm e^{-(1-\eta)\rm s}}{(1+\eta)\rm s}+\frac{4\rm e^{-(1+\eta)\rm s}}{(1-\eta)\rm s}-\frac{8\rm e^{-2\rm s}}{(1-\eta^2)\rm s},\\ \\ \displaystyle 
\left.\mathfrak{f}_{3}\right\vert_{\rm s\to\infty}\simeq-\frac{16\exp(-2\rm s)}{(1-\eta^2)\rm s}+\frac{1}{3\rm s}.
\end{array}
\end{eqnarray} Except for the last term in  $\left.\mathfrak{f}_{1}\right\vert_{\rm s\to\infty}$ and  $\left.\mathfrak{f}_{3}\right\vert_{\rm s\to\infty},$ we may use the  $\exp(-\rm s/b)\simeq1$ so that
\begin{eqnarray}
\begin{array}{c}\displaystyle
\mathrm{G}_{1}^{(\mathrm{H})}\simeq\int_{\rm{T}}^\infty\rm ds \int_{-\eta_0}^{\eta_0}d \eta\left\{\frac{4\rm e^{\rm -(1+\eta) s}}{(1-\eta)\rm s}-\frac{8\rm e^{-2\rm s}}{(1-\eta^2)\rm s}+\frac{\rm e^{\rm -s/b}}{3\rm s}\right\},\\ \\
\displaystyle \mathrm{G}_{2}^{(\mathrm{H})}\simeq4\int_{\rm{T}}^\infty\rm ds \int_{-\eta_0}^{\eta_0}d \eta\left\{\frac{\rm e^{-(1+\eta)\rm s}}{(1-\eta)\rm s}-\frac{\rm e^{-2\rm s}}{(1-\eta^2)\rm s}\right\},\\ \\
\displaystyle \mathrm{G}_{3}^{(\mathrm{H})}\simeq\int_{\rm{T}}^\infty\rm ds \int_{0}^{\eta_0}d \eta\left\{-\frac{16\rm e^{-2\rm s}}{(1-\eta^2)\rm s}+\frac{\rm e^{-\rm s/b}}{3\rm s}\right\}.
\end{array}
\end{eqnarray}
Performing the integration  over $\eta,$
$\mathrm{G}_{i}^{(\mathrm{H})}$ are given by 
\begin{eqnarray}\label{aprodgfu11}
\mathrm{G}_{1}^{(\mathrm{H})}&\simeq&4\int_{\rm{T}}^\infty \rm ds \frac{e^{-2s}}{\rm s}\left[\rm Ei(2s-T)-Ei(T)\right]\\&-&8 \int_{\rm T}^{\infty}\rm ds\ln\left(\frac{2\rm s}{\rm T}-1\right)\frac{\rm e^{-2\rm s}}{\rm s}+\frac{2}{3}\mathrm{Ei}\left(-\frac{\rm T}{\rm b}\right)\nonumber\\&-&\frac{2\rm T}{3}\int_{\rm{T}}^\infty\rm ds\frac{\rm e^{-\rm s/b}}{\rm s^2},\nonumber\\
\mathrm{G}_{2}^{(\mathrm{H})}&\simeq&4\int_{\rm{T}}^\infty \rm ds \frac{e^{-2s}}{\rm s}\left[\rm Ei(2s-T)-Ei(T)\right]\label{aprodgfu22}\\&-&4\int_{\rm{T}}^\infty\rm \frac{ds}{\rm s}\ln\left(\frac{2\rm s}{\rm T}-1\right)\exp(-2\rm s),\nonumber\\
\mathrm{G}_{3}^{(\mathrm{H})}&\simeq&-8\int_{\rm{T}}^\infty\rm \frac{ds}{\rm s}\ln\left(\frac{\rm 2s}{\rm T}-1\right)\rm e^{-2\rm s}+\frac{1}{3}\mathrm{Ei}\left(-\frac{\rm T}{\rm b}\right)\nonumber\\&-&\frac{\rm T}{3}\int_{\rm{T}}^\infty\rm ds\frac{\rm e^{-\rm s/b}}{\rm s^2}\label{aprodgfu33}
\end{eqnarray} where $\rm Ei(-T/b)$ is the exponential-integral function  whose asymptotic expansion for very large  magnetic field   $\rm b\to \infty$  is
\begin{equation}
\int_{\rm T}^{\infty}\frac{\rm ds}{\rm s}\exp\left(-\frac{\rm s}{\rm b }\right)\simeq=\ln\left(\frac{\rm b}{\gamma\pi}\right)-\ln\left(\frac{\rm T}{\pi}\right).\label{eiasyp}
\end{equation}This  expression is  calculated with accuracy of terms that decrease
 with $\rm b$.  To be consistent  the last integral
of Eq. (\ref{aprodgfu11}) and   Eq. (\ref{aprodgfu33}) can  be
neglected as well, since the terms decrease as fast as $\sim\rm
b^{-1}\ln\rm b$ and $\sim\rm b^{-1}.$ The remaining integrals
present in Eq. (\ref{aprodgfu22}) and Eq. (\ref{aprodgfu33}) depend
on the parameter $\rm T.$  Taking all this into account,   the
leading  asymptotic behavior of $\mathrm{G}_{i}$ for $\rm b\to\infty$
reads
\begin{eqnarray}
\begin{array}{c}\displaystyle
\mathrm{G}_{1}\approx\frac{2}{3}\ln\left(\frac{\rm b}{\gamma\pi}\right)+\mathcal{C}_1,\ \ \mathrm{G}_{2}\approx\mathcal{C}_2,\\ \\\displaystyle
\mathrm{G}_{3}\approx\frac{1}{3}\ln\left(\frac{\rm b}{\gamma\pi}\right)+\mathcal{C}_3.\end{array}\label{g3}
\end{eqnarray} where  we have used Eq. (\ref{eiasyp}). Here, the  numerical constants $\mathcal{C}_{1,3}$ are determined by imposing the condition $\rm d \mathcal{C}_{1,3}/dT=0$ with 
\begin{eqnarray}
\mathcal{C}_1&=&4\int_{\rm{T}}^\infty \rm ds \frac{e^{-2s}}{\rm s}\left[\rm Ei(2s-T)-Ei(T)\right]-\frac{19 \rm T^2}{540}\nonumber\\&-&8 \int_{\rm T}^{\infty}\rm ds\ln\left(\frac{2\rm s}{\rm T}-1\right)\frac{\rm e^{-2\rm s}}{\rm s}-\frac{2}{3}\ln\left(\frac{\rm T}{\pi}\right),\\
\mathcal{C}_3&=&-8\int_{\rm{T}}^\infty\rm \frac{ds}{\rm s}\ln\left(\frac{\rm 2s}{\rm T}-1\right)\rm e^{-2\rm s}-\frac{1}{3}\ln\left(\frac{\rm T}{\pi}\right)\nonumber\\&-&\frac{41\rm T^2}{1620}.
\end{eqnarray} This yields 
\begin{equation}
\mathcal{C}_1\approx2.67,\ \ \rm and\ \ \mathcal{C}_3\approx0.18.\label{numericalcosntaudcna}
\end{equation} Note that there is not value  fulfilling  the condition $\rm d \mathcal{C}_{2}/dT=0.$ However, in order to compute it, we first set  $\frac{\rm d}{\rm dT}\sum_{i=1}^3\rm G_i=0.$  This condition  leads to $\rm T\simeq0.46$ and
\begin{equation}
\rm G=\sum_{i=1}^3G_i\approx\ln\left(\frac{\rm b}{\gamma\pi}\right)-1.82.
\end{equation} Obviously, $\rm G_2=G-G_1-G_3.$ Thus, by taking into
account  Eqs. (\ref{g3}) and Eq. (\ref{numericalcosntaudcna}) we
obtain $ \mathcal{C}_2\approx-4.68.$ Such that
\begin{eqnarray}
\begin{array}{c}\displaystyle
\mathrm{G}_{1}\approx\frac{2}{3}\ln\left(\frac{\rm b}{\gamma\pi}\right)+2.67,\ \ \mathrm{G}_{2}\approx-4.68,\\ \\ \displaystyle
\mathrm{G}_{3}\approx\frac{1}{3}\ln\left(\frac{\rm b}{\gamma\pi}\right)+0.18.\end{array}\label{g3v}
\end{eqnarray} Substitution of  Eqs. (\ref{g3v}) in Eq. (\ref{secondpartoftwoloopterm}) allows to obtain
\begin{eqnarray}
\mathfrak{L}_{1\mathrm{R}}^{(2)2}&\approx&-\frac{\alpha\rm m^4\rm b^2}{16 \pi^3}\left[\frac{1}{3}\ln\left(\frac{\rm b}{\gamma\pi}\right)+1.34\right],\label{secondpart2loop1}\\
\mathfrak{L}_{2\mathrm{R}}^{(2)2}&\approx&-\frac{\alpha\rm m^4\rm b^2}{32 \pi^3}\mathcal{C}_2\ \ \mathrm{with}\ \ \mathcal{C}_2=-4.68,\\ \mathfrak{L}_{3\mathrm{R}}^{(2)2}&\approx&-\frac{\alpha\rm m^4\rm b^2}{32 \pi^3}\left[\frac{1}{3}\ln\left(\frac{\rm b}{\gamma\pi}\right)+0.18\right].\label{secondpart2loop2}
\end{eqnarray}Finally, the  behavior of $\mathfrak{L}_{i\mathrm{R}}^{(2)}$ presented in Eq. (\ref{leading1})
is derived by inserting the expression bellow Eqs. (\ref{firstpart2loop11}) and  Eqs. (\ref{secondpart2loop1}-\ref{secondpart2loop2})
into   Eq. (\ref{sppliting}).

\end{document}